\documentclass{article}

\usepackage[utf8]{inputenc}
\usepackage{graphicx}
\usepackage{amsmath}
\usepackage{booktabs}
\usepackage{caption,subcaption}
\usepackage[export]{adjustbox}
\usepackage{soul}
\usepackage{comment}
\usepackage{multirow}
\usepackage{chngcntr}
\usepackage{xcolor}
\usepackage{authblk}
\usepackage{tikz}
\usetikzlibrary{shapes.geometric, arrows}
\usepackage[title]{appendix}
\usepackage[left=2cm, right=2cm, top=2cm]{geometry}
\usepackage{framed}
\usepackage{upgreek}
\usepackage{amssymb}
\usepackage{notes2bib}
\usepackage{array}
\newcolumntype{P}[1]{>{\centering\arraybackslash}p{#1}}

\newcommand{\nohl}[1]{{#1}}

\tikzstyle{process} = [rectangle, minimum width=3cm, minimum height=1cm, text centered, draw=black, fill=gray!15]
\tikzstyle{decision} = [diamond, minimum width=2cm, minimum height=0.5cm, text centered, draw=black, fill=gray!15]
\tikzstyle{arrow} = [thick,->,>=stealth]
\tikzset{
  |-/.style={to path={|- (\tikztotarget) \tikztonodes}},
  -|/.style={to path={-| (\tikztotarget) \tikztonodes}} }

\title{Dislocation Density-Based Plasticity Model from Massive Discrete Dislocation Dynamics Database}

\author[1]{Sh. Akhondzadeh}
\affil[1]{Department of Mechanical Engineering, Stanford University, Stanford CA, 94305, USA}
\author[2]{Ryan B. Sills}
\affil[2]{Department of Materials Science and Engineering, Rutgers University, Piscataway, NJ, 08854, USA}
\author[1,3]{Nicolas Bertin}
\affil[3]{Lawrence Livermore National Laboratory, Livermore, CA, 94550, USA}
\author[1]{Wei Cai}
\date{\today}

\begin{document}
\maketitle

\begin{abstract}
We present a dislocation density-based strain hardening model for single crystal copper through a systematic coarse-graining analysis of more than 200 discrete dislocation dynamics (DDD) simulations of plastic deformation under uniaxial tension.
The proposed constitutive model has two components: a generalized Taylor relation connecting resolved shear stresses to dislocation densities on individual slip systems, and a generalized Kocks-Mecking model for dislocation multiplication.
The DDD data strongly suggests a logarithmic dependence of flow stress on the plastic shear strain rate on each slip system, and, equivalently, an exponential dependence of the plastic shear strain rate on the resolved shear stress.  Hence the proposed generalized Taylor relation subsumes the Orowan relation for plastic flow.
The DDD data also calls for a correction to the Kocks-Mecking model of dislocation multiplication to account for the increase of dislocation density on slip systems with negligible plastic shear strain rate.  This is accomplished by allowing the multiplication rate on each slip system to include contributions from the plastic strain rates of the two coplanar slip systems.
The resulting constitutive model successfully captures the strain hardening rate dependence on the loading orientation as predicted by the DDD simulations, which is also consistent with existing experiments.
\end{abstract}

\section{Introduction}
\label{sec:intro}

Most metals are crystalline materials that can undergo significant plastic deformation before \nohl{fracture}. This plastic deformation is usually accompanied by an increase in the flow stress of the material, a phenomenon which is called strain hardening and is of vital importance in many engineering applications, including aerospace, automotive, and power generation industries.
While most metals used in engineering applications come in the form of polycrystalline aggregates, a comprehensive understanding of the behavior of their individual constituents, i.e., single crystals, is fundamental to understand and develop accurate predictive capabilities for the plastic  response and hardening behavior of metals during deformation.

In single crystals, slip induced by dislocation motion is the dominant mechanism for plastic deformation under most conditions. In this case the flow stress of the crystal is governed by the evolution of dislocations moving and interacting in response to the applied loading.
Thus, following the motion of individual dislocations at the micron-scale can provide the physical link between the dislocation microstructure evolution and the strain hardening phenomenon. 
To this end, Discrete Dislocation Dynamics (DDD) simulations have been developed to predict the plastic flow response of crystals from the collective motion of individual dislocation lines~\cite{bulatov2006computer}.
However, the high computational cost of DDD simulations limits the \nohl{accessible} length and time scales of configurations with realistic initial dislocation densities to $\sim10\,\upmu \rm m$ simulation sizes and $\sim10\,\upmu \rm s$ of simulated times~\cite{arsenlis2007enabling,rao2008athermal, ARMRnbertin}. These scales are still well below those which are relevant to most engineering applications.

To enable predictions of mechanical response at the macroscale for engineering applications, a large number of plasticity models suitable for crystal plasticity (CP) \nohl{simulations} have been proposed~\cite{roters2010overview}.
In contrast to lower scale models such as DDD, CP models require constitutive relations to describe plastic flow response as a function of the material state.
An example of these is the widely used flow rule that relates the plastic strain rate $\dot{\gamma}_i$ on slip system $i$ to the resolved shear stress $\tau_i$ through a power-law~\cite{hutchinson1976bounds,peirce1983material,roters2010overview}:
\begin{equation}
    \dot{\gamma}_i = \dot{\gamma}_0 \left( \frac{\tau_i}{\tau_i^{\rm c}} \right)^{1/m}
    \label{eq:powerlaw}
\end{equation}
where $\dot{\gamma}_0$ is a reference strain rate, $m$ is an exponent characterizing the rate sensitivity of the model, and $\{\tau_i^{\rm c}\}$, referred to as the critical resolved shear stresses, is a set of microstructural variables describing the resistance to dislocation slip on systems $i = 1, \cdots, N_{\rm sys}$ where $N_{\rm sys}$ is the total number of slip systems.
To date, the most outstanding challenge in the framework of CP models is that Eq.~\eqref{eq:powerlaw} and other constitutive relations~\cite{franccois2012mechanical,kocks1976laws,becker1991analysis,johnson1983constitutive} commonly used in engineering calculations~\cite{roters2019damask} are still phenomenological, i.e., \nohl{they are not derived from fundamental dislocations physics}~\cite{MSMSE2020}.
These models are typically parametrized by fitting to experimental data under uniaxial loading~\cite{ma2004constitutive,groh2009multiscale,chandra2015multiscale}.
As a result, their extensibility to conditions beyond those used for fitting as well as their predictive capabilities under more complex scenarios remains unclear.
Furthermore, numerous researchers have demonstrated that the local strain distributions predicted by the existing CP models do not compare well with experimental measurements~\cite{pinna2015assessment,mello2016effect}.

In this work, we take a different approach to constitutive model development and demonstrate how constitutive models of strain hardening can be constructed from lower-scale, physics-based dislocation models through a systematic coarse-graining analysis of a large DDD simulation database.
Thanks to recent advances in DDD, including the subcycling time-integration algorithm~\cite{sills2016advanced} and its implementation on Graphics Processing Units (GPUs) \cite{bertin2019gpu}, we can now perform large DDD simulations on representative ensembles of dislocations for periods of time long enough that consistent hardening rates can be extracted (in the regime of small plastic shear strain $\gamma < 5\%$).
Here we take advantage of this new capability to generate a large DDD database of more than 200 simulations in FCC Cu with uniaxial loading orientations \nohl{which cover the} stereographic triangle. For each simulation dislocation density, plastic shear strain rate, and other relevant microstructural features on every slip system were extracted as functions of time.
A recent study~\cite{stricker2015dislocation} also made use of a large number of DDD simulations to find correlations between simulation features. Their findings were used to aid development of continuum dislocation dynamics (CDD) models~\cite{sudmanns2019dislocation, roters2019damask}, \nohl{which are models at a smaller scale than CP}.
\nohl{Here we use DDD data to directly extract constitutive relations suitable for CP.}
As we will show, analyzing DDD data not only allows us to test the validity of existing constitutive relations \nohl{in CP}, but also provides important feedback that guides us in an iterative, trial-and-error process to improve the constitutive model.

While \nohl{an} essential constitutive function employed in CP modeling is the flow rule dictating the rate of plastic flow (e.g., Eq.~(\ref{eq:powerlaw})), we show below that in coarse-graining our DDD data it is instructive to focus instead on the flow stress.
We can then utilize our understanding of the flow stress to obtain a flow rule.
For instance, rewriting Eq.~\eqref{eq:powerlaw} it immediately follows that the flow stress on slip system $i$ is
\begin{equation}
    \tau_i = \tau_i^{\rm c} \left( \frac{\dot{\gamma}_i}{\dot{\gamma}_0} \right)^{m}.
    \label{eq:invpowerlaw}
\end{equation}
This can be interpreted in terms of a strain-rate-independent and a strain-rate-dependent contribution to the flow stress (hereafter referred to in terms of ``rate-dependence'').
We can express Eq.~(\ref{eq:invpowerlaw}) as $\tau_i = f^{\rm ri}(\tau_i^{\rm c})f^{\rm rd}(\dot{\gamma}_i)$, where $f^{\rm ri}(\cdot)$ and $f^{\rm rd}(\cdot)$ are functions describing the rate-independent and rate-dependent contributions to the flow stress, respectively.
Hence, the flow rule in Eq.~(\ref{eq:powerlaw}) implies a ``multiplicative decomposition'' of the flow stress in Eq.~(\ref{eq:invpowerlaw}).
The validity of such an assumption needs to be assessed using a lower scale model.
Here we will show how DDD simulations can provide insights into the form of the constitutive equations, and will show that, for the high strain rates considered here, the rate dependent and independent contributions to the flow stress follow an additive decomposition instead. 

The task of developing a constitutive model of crystal plasticity through coarse-graining of discrete defects simulations can be conceptually broken down into two major steps: (i) selection of the appropriate state variables to obtain a physically relevant coarse-grained description of the dislocation microstructure, and (ii) identification of relations describing the time evolution of these state variables in response to the applied loading and current stress state.
The first step is important for the success of the second. 
The state variables need to be carefully chosen so that a closed-set of relations can be found that express the evolution rates solely in terms of the chosen set of variables themselves and the current stress state and temperature.
\nohl{In contrast to} Eq.~\eqref{eq:invpowerlaw}, some CP models express the flow stress in terms of various dislocation density-based microstructural variables, including the total dislocation density ($\rho$)~\cite{kocks2003physics}, dislocation density per slip system ($\rho_i$)~\cite{devincre2008dislocation}, mobile and immobile dislocation densities per slip system~\cite{ma2004constitutive}, forest dislocation density per slip system~\cite{demir2016physically}, and dislocation density per slip system including contributions from junctions~\cite{kubin2008modeling}.
Clearly, the larger the set of microstructural variables, the better agreement with the DDD data is likely to be achieved.
However, a larger set of microstructural variables also requires a larger set of  parameters to describe their interactions in a more complex set of evolution equations. 
At the same time, constructing a model with too many parameters based on a limited set of data has the danger of over-fitting.

In this work, our goal is to develop a constitutive model for plasticity of FCC single crystals. 
We specifically seek a constitutive model that satisfies the following two criteria: (1) be as simple as possible, and (2) be consistent with the DDD data and the dislocation behaviors exhibited in the DDD simulations.
To some extent, these two criteria are in conflict with each other, because a simple model with too few state variables may not have enough flexibility to match the DDD simulation data well. 
Here, according to criterion (1), we found it sufficient to construct a model based on the dislocation densities, $\rho_i$, on each of the 12 slip systems of the FCC crystal.
We show that with appropriate adjustments to the Taylor and Kocks-Mecking relations which describe flow stress and dislocation multiplicaton rates, respectively, this dislocation density-based model with 12 degrees of freedom is able to capture the dislocation density time-history for each slip system and the overall strain hardening rate reasonably well when compared with DDD simulations for over a hundred loading orientations.

The remainder of the paper is organized as follows. First the DDD simulation setup and the procedure of microstructure data extraction are described in Section~\ref{sec:DDDsetup}. 
The strain hardening rates predicted by these DDD simulations as a function of loading orientations in the stereographic triangle are presented in Section~\ref{sec:HRates}.  Good qualitative agreement is observed with existing experimental measurements.
In Section~\ref{sec:modfTaylor}, the correlation between the resolved shear stress on the dominant slip system and the dislocation densities is investigated to reveal the rate-independent and rate-dependent flow stress contributions. 
The DDD data suggests  that the generalized Taylor law describes the rate-independent part, with the rate-dependent contribution exhibiting a logarithmic dependence on the plastic strain rate on the dominant slip system.
In Section~\ref{sec:Taylor-link} this expression is justified by examining the physical origin of the Taylor law.
We find that an even simpler form for the flow stress---requiring far fewer parameters---can be \nohl{obtained} in terms of the average dislocation link length (e.g., length of dislocation line between junction nodes).
Using this expression, in Section~\ref{sec:exponentialEqVelocity} we obtain the flow rule implied by the DDD data.
In Section~\ref{sec:rhoidot}, we construct an evolution equation for the dislocation multiplication rate, $\dot{\rho}_i$.  The DDD data shows that the Kocks-Mecking model needs to be modified to account for non-zero $\dot{\rho}_i$ on slip systems where $\dot{\gamma}_i \approx 0$, \nohl{a phenomenon we refer to as \emph{slip-free multiplication}}.
This is accomplished by allowing the multiplication rate on slip system $i$ to include contributions from plastic shear strain rates on the two slip systems which are coplanar with slip system $i$, motivated by the role of coplanar dislocation interactions identified in the DDD simulations.
In Section~\ref{sec:fullModel} we show that the proposed constitutive equations are able to reproduce strain hardening rates consistent with DDD predictions. 
A discussion on the implications and limitations of our results is given in Section~\ref{sec:discussion}.

\section{Methods} 
\label{sec:DDDsetup}

The DDD simulations were performed using the ParaDiS program~\cite{arsenlis2007enabling} with the recently developed subcycling time integration algorithm~\cite{sills2016advanced} and its GPU implementation~\cite{bertin2019gpu}.
Material properties for copper are used, with shear modulus $\mu = 54.6$~GPa, Poisson's ratio $\nu = 0.324$, and Burgers vector magnitude $b = 0.255$~nm. Glissile dislocations on the $\frac{1}{2}\langle110\rangle\{111\}$ slip systems follow a linear mobility law with drag coefficient $B = 1.56\times 10^{-5}$~Pa$\cdot$s. Dislocation junctions are only allowed to move along their own line direction by the zipping/unzipping mechanism.
Cross-slip was not allowed in any of the simulations (see Section~\ref{sec:crossslip}).

\begin{figure}[ht]
\begin{subfigure}[t]{0.49\textwidth} 
 \includegraphics[scale=0.18,center]{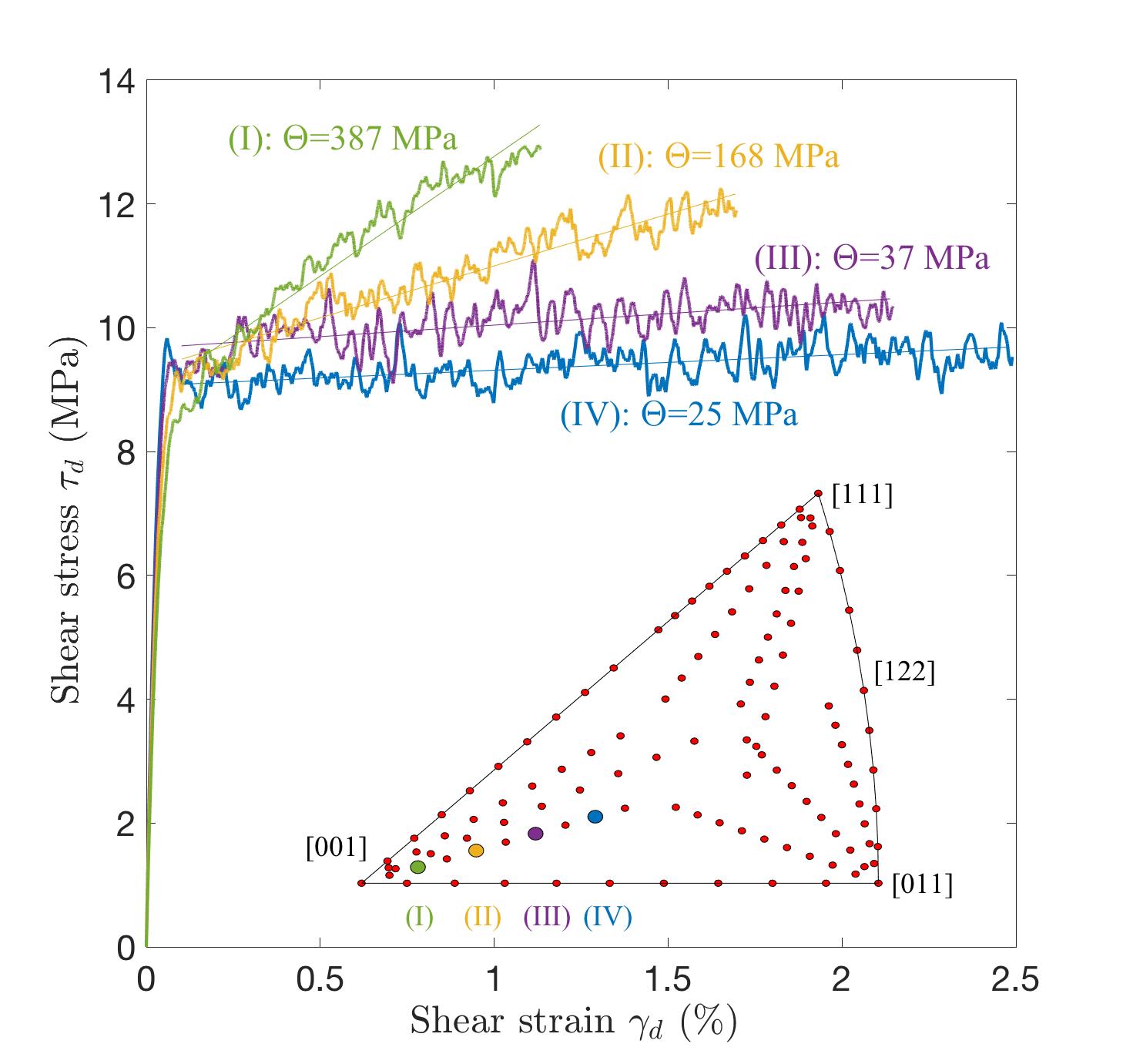}
 \caption{}
 \label{fig:SigmaEps_SigmaEps}
 \end{subfigure}
 \begin{subfigure}[t]{0.49\textwidth}
 \includegraphics[scale=0.18,center]{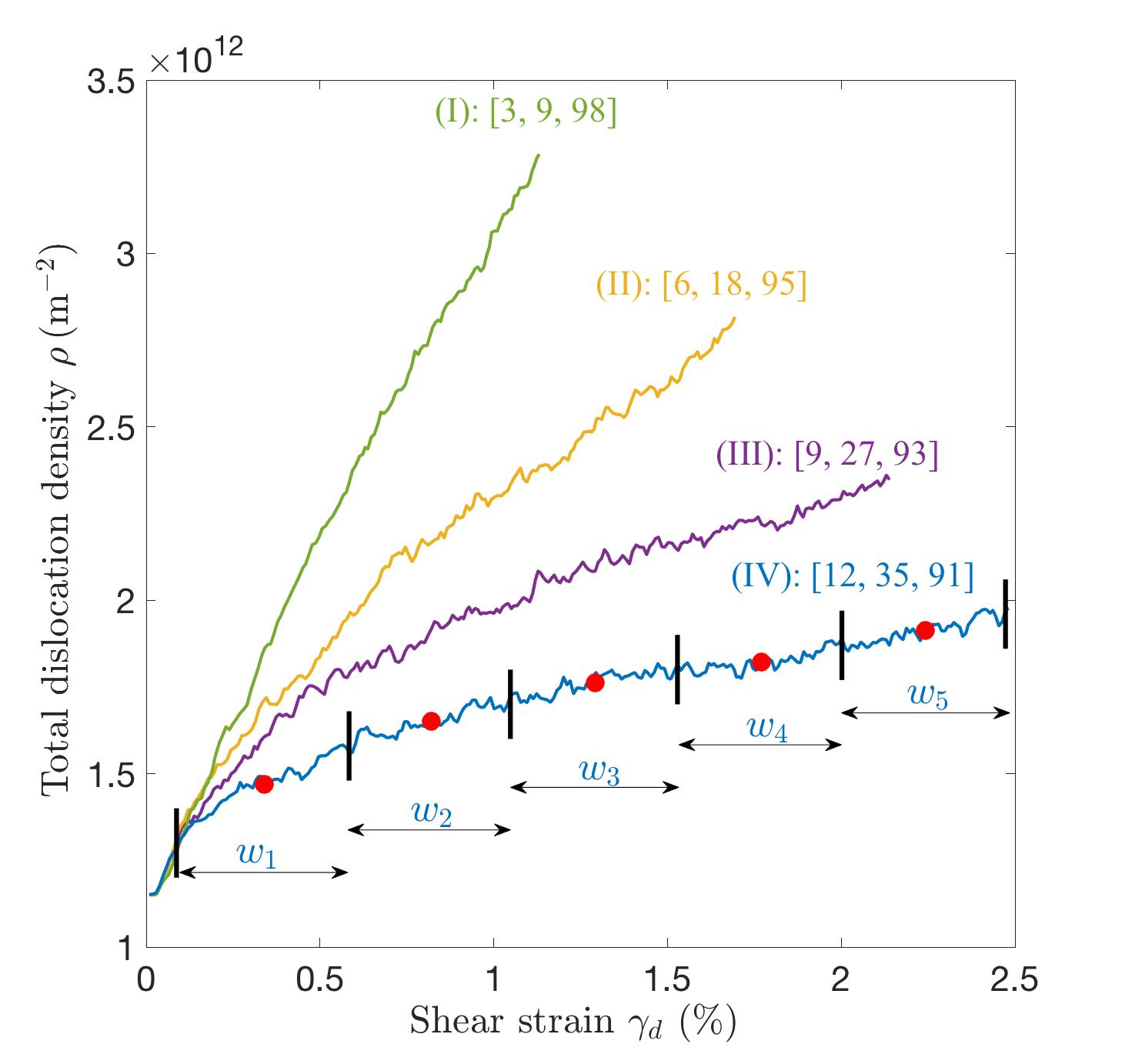}
 \caption{}
 \label{fig:SigmaEps_Density}
 \end{subfigure}
 \caption{(a) Four example shear stress-shear strain curves predicted by DDD simulations, along the specified loading orientation, subjected to $\dot{\varepsilon}=10^3 \, {\rm s^{-1}}$. $\Theta$ denotes the hardening rate. \nohl{The inset shows the loading orientation associated with each of the 120 simulations mapped to the stereographic projection triangle.} (b) \nohl{Total dislocation density as a function of shear strain for the four loading orientations shown in (a).}  The windows of averaging, denoted by $w_r$, $r=1,2,..,5$ are illustrated on curve IV and averaged values are shown by red dots.
 }
\label{fig:SigmaEps}
\end{figure}

The initial configuration was generated by randomly introducing straight dislocation lines into a $(15\,{\rm \mu m})^3$ cell subjected to periodic boundary conditions in all three directions.  After relaxation the initial dislocation density is $\rho_0 \approx 1.2\times 10^{12}\, {\rm m^{-2}}$.
The relaxed configuration is then subjected to uniaxial tension at a constant strain rate of $\dot{\varepsilon} = 10^3\,{\rm s}^{-1}$ along 120 different directions, sampled in the symmetry-irreducible stereographic triangle, as shown in the inset of figure~\ref{fig:SigmaEps}a.
The simulations were performed until the shear strain on the dominant slip system reaches the values in the range of $\gamma_d \approx$ $1\%$ to $3\%$.
Examples of the shear stress \nohl{v.s.} shear strain curves corresponding to four different loading orientations are shown in figure~\ref{fig:SigmaEps}a, and the corresponding evolutions of the total dislocation density as a function of strain are shown in figure~\ref{fig:SigmaEps}b.
Unless stated otherwise, the DDD results discussed in this paper refer to simulations under the strain rate of $\dot{\varepsilon} = 10^3\,{\rm s}^{-1}$.
To examine the effect of strain rate, a smaller set of DDD simulations are performed under strain rates of $\dot{\varepsilon} = 10^2\,{\rm s}^{-1}$ and $ 10^4\,{\rm s}^{-1}$, for 27 and 9 different loading orientations, respectively. In order to verify that our findings do not depend on the initial dislocation configuration, we also repeated our strain hardening simulations using a second random initial dislocation configuration, under the strain rate of $\dot{\varepsilon} = 10^3\,{\rm s}^{-1}$ along 54 different loading orientations, and under $\dot{\varepsilon} = 10^2\,{\rm s}^{-1}$ and  $10^4\,{\rm s}^{-1}$ along 27 and 9 different orientations, respectively.

During the simulations, we recorded, in addition to macroscopic quantities such as stress and strain, the evolution of relevant microstructural parameters at the slip system level, including dislocation densities $\rho_i$, the accumulated plastic shear $\gamma_i$, average dislocation link lengths $\bar{l}_i$, etc. 
Due to the statistical nature of DDD simulations, each of these variables fluctuates during the simulations.
To help construct a continuum model of strain hardening in which the state variables evolve smoothly with time, some averaging is thus needed to reduce the fluctuations in the raw DDD data.
To this end, the raw data trajectory of each simulation is divided into a specified number of blocks labelled as $w_r$, with $r=1,...,N_w$ and $N_w=5$ or 9 (see figure \ref{fig:SigmaEps_Density}), and the time-averaged values for $\rho_i$ and $\bar{l}_i$ are computed for each block and assigned to the time center of the block. A similar averaging procedure was used in~\cite{csikor2005numerical}.
To compute the plastic strain rate $\dot{\gamma}_i$ for each slip system, we first fit the raw ($\gamma_i$, $t$) data to a third order polynomial of time $t$.  We then take an analytic derivative of the polynomial and evaluate $\dot{\gamma}_i$ at the center time of each block.
Unfortunately, the above procedure cannot be applied \nohl{reliably} to evaluate the dislocation multiplication rate $\dot{\rho}_i$, because of large fluctuations of $\rho_i(t)$.
Therefore, we do not evaluate $\dot{\rho}_i$ from the DDD data.  Instead, we numerically integrate the $\dot{\rho}_i$ expression in the \nohl{coarse-grained} model and compare it against the averaged $\rho_i$ values from DDD data.
\nohl{In order to ensure robust conclusions when quantifying the goodness-of-fit between a model and our DDD database, we have utilized two different error measures: root mean square error (RMSE), and the coefficient of determination $R^2$.}
%

\section{Results} 
\label{Microstructure_analysis}

\subsection{Hardening rates}
\label{sec:HRates}

\begin{figure}[ht]
    \centering
    \includegraphics[trim={0 0 0 14cm},clip=true,scale=0.13]{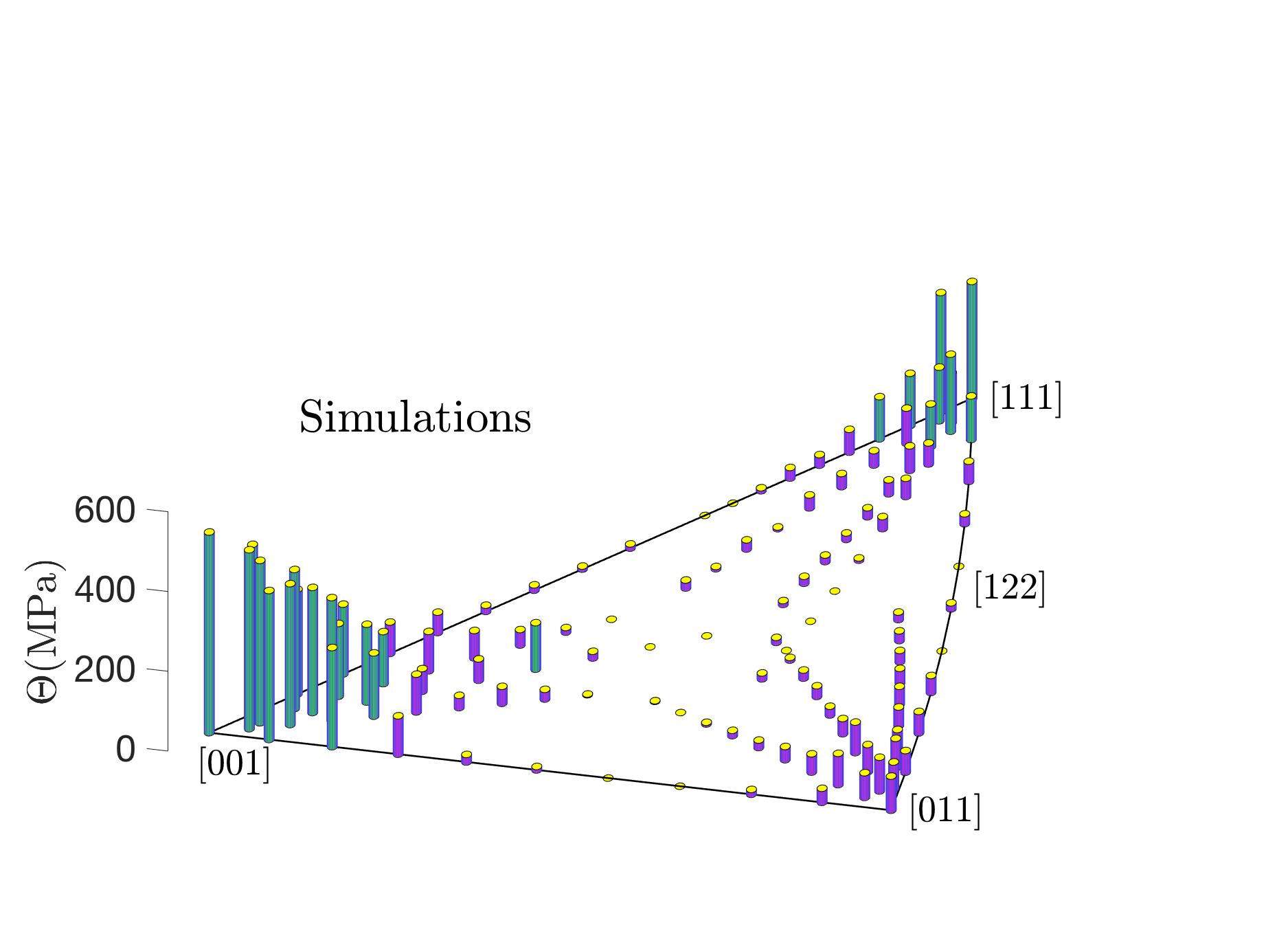}
    \includegraphics[trim={0 0 0 14cm},clip=true,scale=0.13]{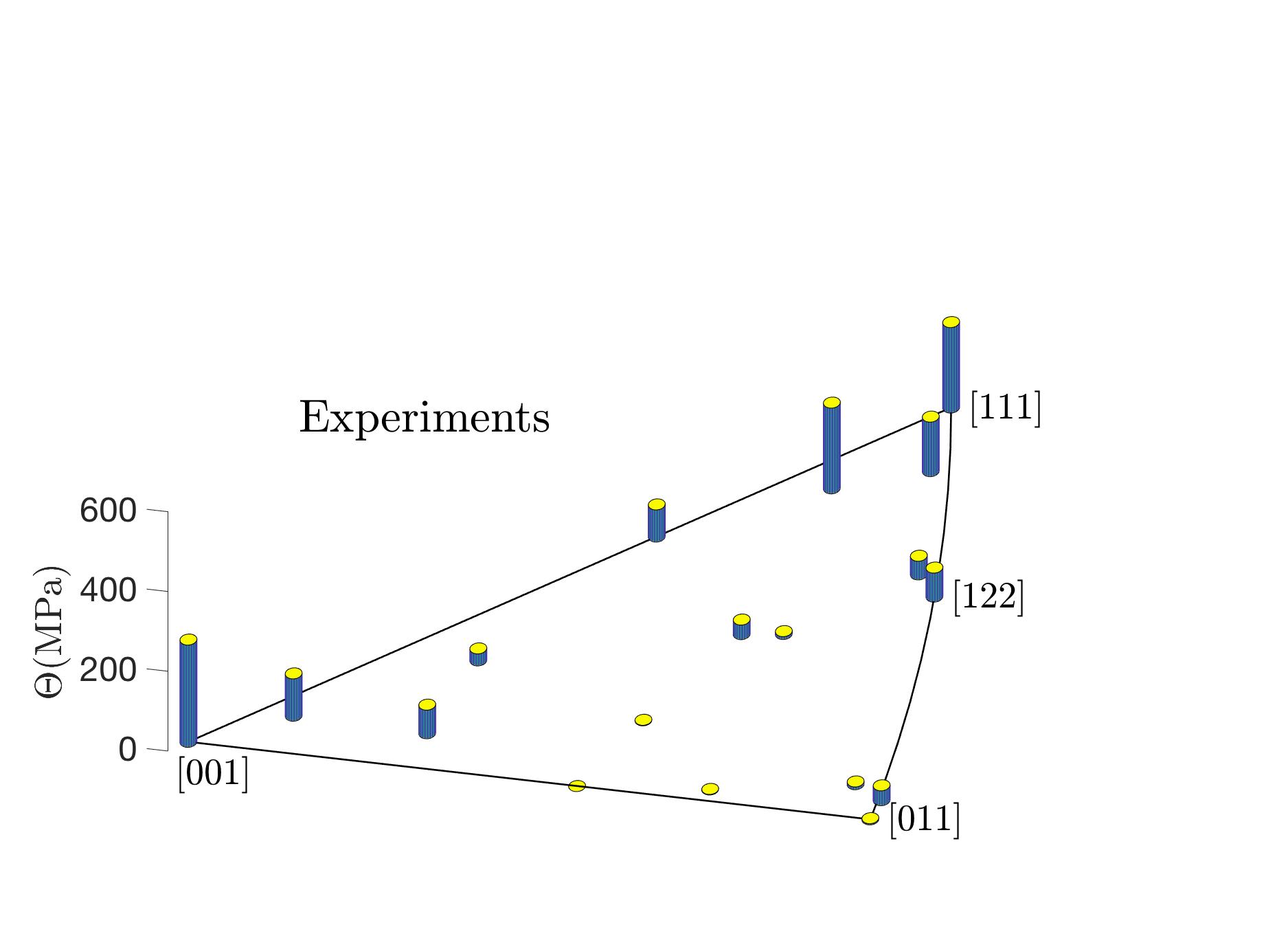} \\
    (a) \hspace{2.6in} (b)
    \caption{Strain hardening rates $\Theta$ of single crystal copper as a function of loading orientation, represented as bars in the stereographic triangle. (a) Strain hardening rates from DDD simulations under constant strain rate $\dot{\varepsilon}=10^3\,{\rm s^{-1}}$ along 120 different loading directions. High-hardening and low-hardening loading orientations are shown by different colors. (b) Strain hardening rates extracted from experimental stress-strain curves~\cite{honeycomb1972plastic, takeuchi1975work} up to shear strain of $\gamma=5\%$ under quasi-static strain rate of $\dot{\varepsilon}\approx 3\times10^{-3} \, {\rm s^{-1}}$.}
\label{fig:stTri}
\end{figure}

From the shear stress-strain curve predicted by the DDD simulation for each loading orientation, the strain hardening rate $\Theta \equiv d\tau/d\gamma$ is extracted by fitting a straight line to the post-yield regime, as shown in figure~\ref{fig:SigmaEps_SigmaEps}.
Figure~\ref{fig:stTri}a shows the resulting strain hardening rates for 120 different loading orientations sampled in the stereographic triangle.
Note that since our simulations reach up to a few percent shear strain, crystal rotations are very small. Hence, the difference between the initial and final loading orientations is ignored in the subsequent discussions and for simplicity we ignore the slight variations of the Schmid factors, $S_i$, during the deformation.

From figure~\ref{fig:stTri}a it can be seen that loading orientations in the regions near the $[001]$ and $[111]$ orientations correspond to high strain hardening \nohl{rates}.  On the other hand, there is very little hardening for loading orientations near the $[011]$ direction and those near the center of the stereographic triangle.
The predicted orientation dependence of the strain hardening rate is in good qualitative agreement with experimental values shown in figure~\ref{fig:stTri}b.  The experimental strain hardening rates are extracted from shear stress-strain curves up to $\gamma_d=5\%$ of single crystal copper~\cite{honeycomb1972plastic,takeuchi1975work} deformed under a quasi-static strain rate of $\dot{\varepsilon} \approx 3 \times 10^{-3}\, {\rm s}^{-1}$ at room temperature. 
For example, the strain hardening rates along $[001], [111]$ and $[011]$ predicted by DDD simulations are 508, 293 and 85~MPa, respectively, while the corresponding experimental values are 256, 214, 10 MPa \cite{takeuchi1975work}, respectively.
Hence our DDD simulations successfully capture the order of the strain hardening rates in these three orientations.
The discrepancies in the quantitative values of the strain hardening rate may be attributed to the differences in the applied strain rate.

\nohl{In the analysis below we label one of the active slip systems in each simulation as the \emph{dominant} slip system, designated by index $d$, which we define as the system for which the Schmid factor $S_d$ is the largest.
Under highly symmetric loading conditions, multiple slip systems may have the same maximum Schmid factor, in which case, the slip system with the highest plastic strain rate $\dot{\gamma}_d$ is taken as the dominant one.}
Loading orientations can be categorized into two groups based on the magnitude of their hardening rates.
In the literature, orientations with high and low strain hardening rates are often associated with multi-slip (multiple slip systems are active) and single-slip (one slip system is active) orientations, respectively. 
The rationale is that a higher number of dislocation intersections occur when multiple slip systems are simultaneously active, resulting in a higher hardening rate. 
However, we find that there \nohl{exist} loading orientations which exhibit a low hardening rate while more than one (and up to four) slip systems are active (e.g., $[110]$).
\nohl{Here a slip system $i$ is considered as \emph{active} when its plastic strain rate is at least 10\% of the plastic strain rate of the dominant slip system, i.e., when  $\dot{\gamma}_i > 0.1\times \dot{\gamma}_d$.}
Therefore, to prevent terminological ambiguity, in this work we will refer to \emph{low-hardening} ($\Theta < 100$~MPa) and \emph{high-hardening} ($\Theta \ge 100$~MPa) orientations, rather than employing the usual single-slip / multiple-slip  distinction.
\nohl{Out of the 174 simulations at $\dot{\varepsilon}=10^3\,s^{-1}$ analyzed, 42 of them were categorized as high-hardening; the remaining 132 are characterized as low-hardening.}

\subsection{A rate-dependent Taylor law}
\label{sec:modfTaylor}

As a first step in developing a constitutive model for strain hardening based on the DDD data, we examine the relationship between the flow stress and the dislocation density.
The Taylor law states that the flow stress is proportional to the square root of the dislocation density~\cite{taylor1934mechanism}
\begin{equation}
  \tau_d=\alpha \mu b \sqrt{\rho} + \tau^{\rm rd}
  \label{Eq:Taylor}
\end{equation}
where $\tau_d$ denotes the resolved shear stress on the dominant slip system $d$ and $\alpha$ is a constant, which for pure Cu is estimated to be in range $0.2-0.5$~\cite{neuhaus1992flow, madec2002dislocation}.
Since the Taylor law was initially established for quasi-static loading conditions (e.g., rate-independent flow stress)~\cite{taylor1934mechanism,mecking1981kinetics,neuhaus1992flow} and the flow stress is known to increase at high strain rate conditions \cite{edington1969influence, akhondzadeh2018geometrically}, we have added a (rate-dependent) correction term $\tau^{\rm rd}$ to Eq.~(\ref{Eq:Taylor}) to account for the high strain rates here.
Since all of the simulations analyzed in this section correspond to the same applied strain rate ($10^3\,\rm s^{-1}$), we initially treat the rate-dependent correction as a constant.

\begin{figure}[ht]
        \begin{subfigure}[t]{0.3\textwidth} 
        \includegraphics[trim={0 0 3cm 0},clip=true,scale=0.14,right]{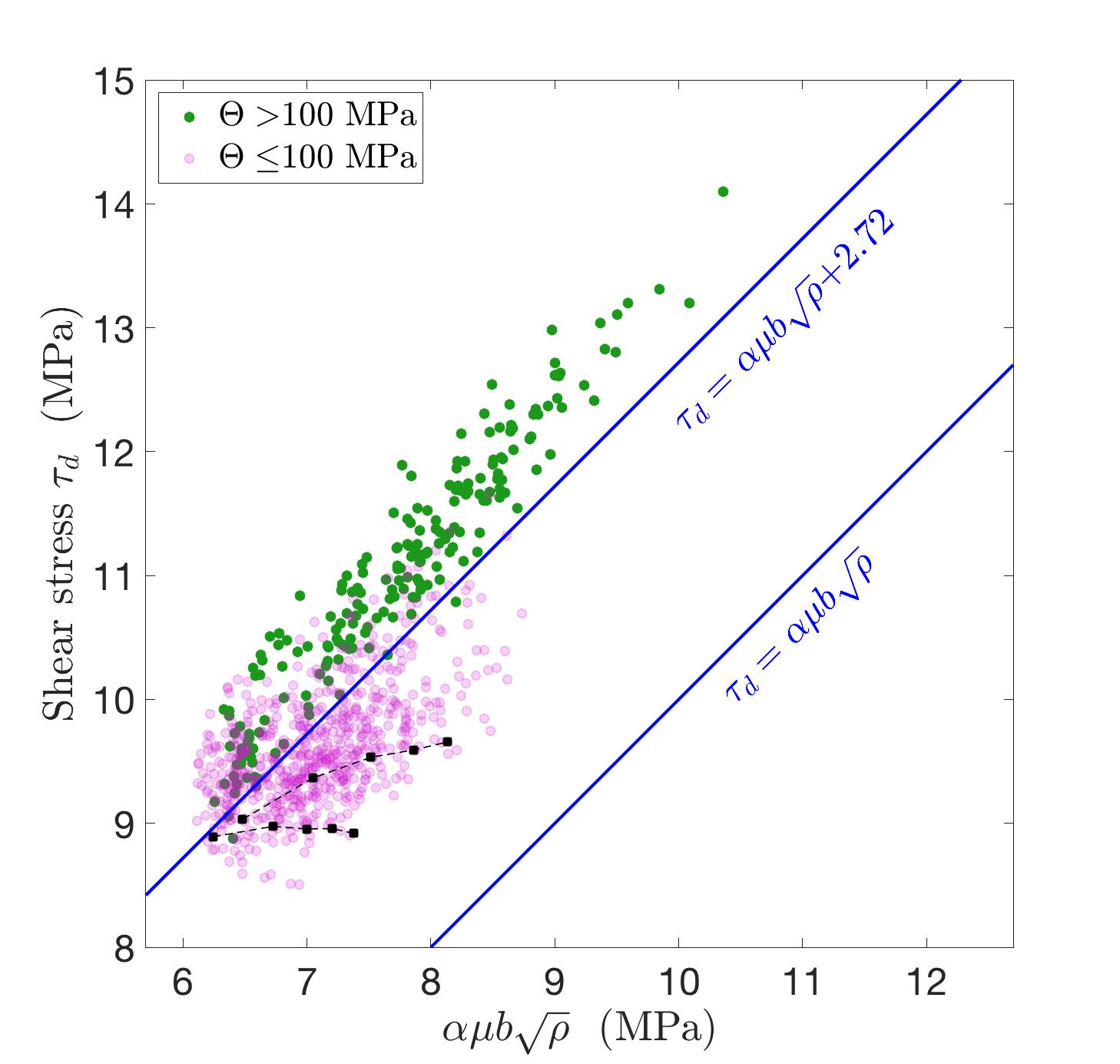}
        \caption{}
        \label{fig:Tau_totalRho}
    \end{subfigure}\hfill
    \begin{subfigure}[t]{0.3\textwidth} 
        \includegraphics[trim={4.5cm 0 0 0cm},clip=true,scale=0.14,right]{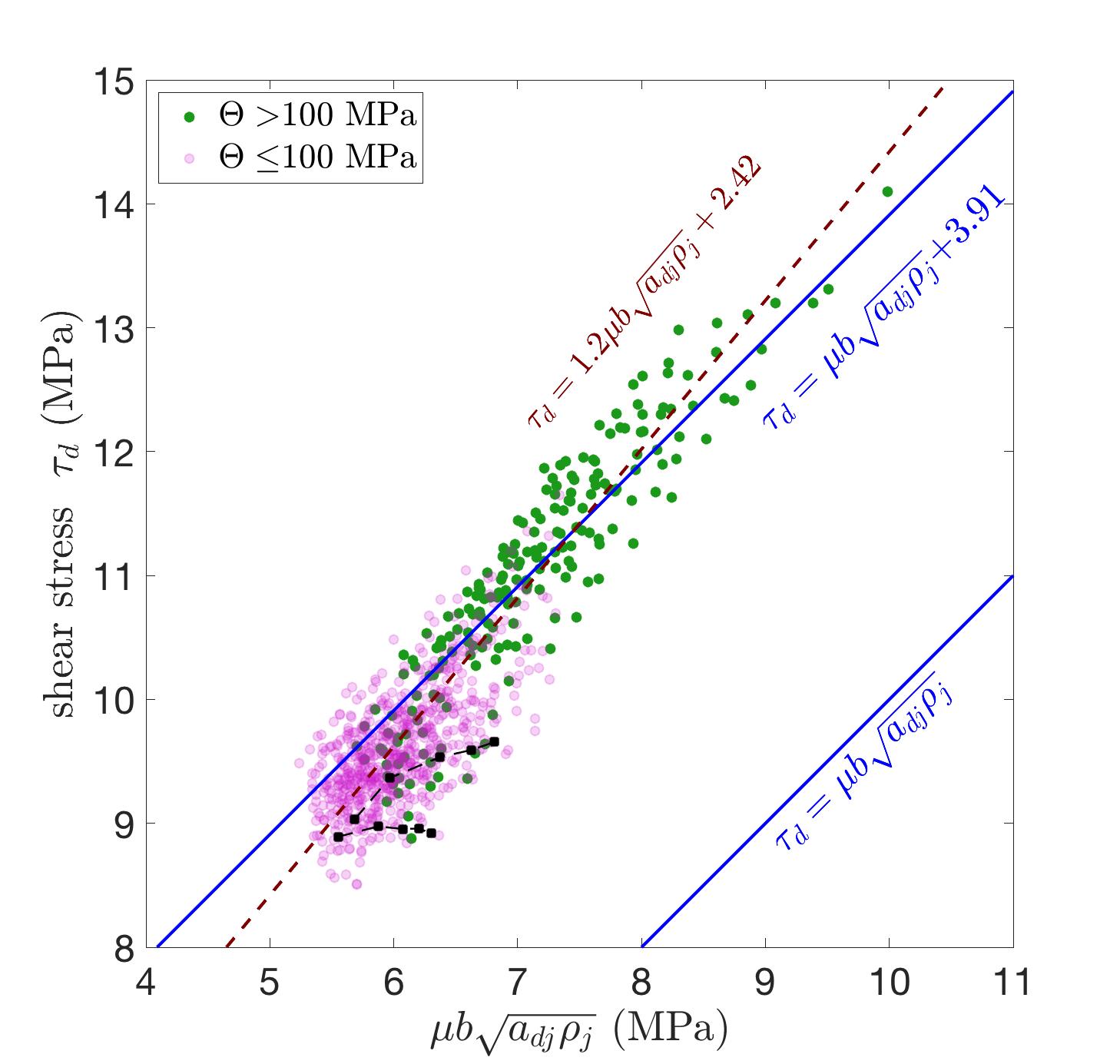}
        \caption{}
        \label{fig:Tau_aijRhoj_Kubin}
    \end{subfigure}
    \begin{subfigure}[t]{0.3\textwidth} 
        \includegraphics[trim={4.5cm 0.2cm 0 0},clip=true,scale=0.14,left]{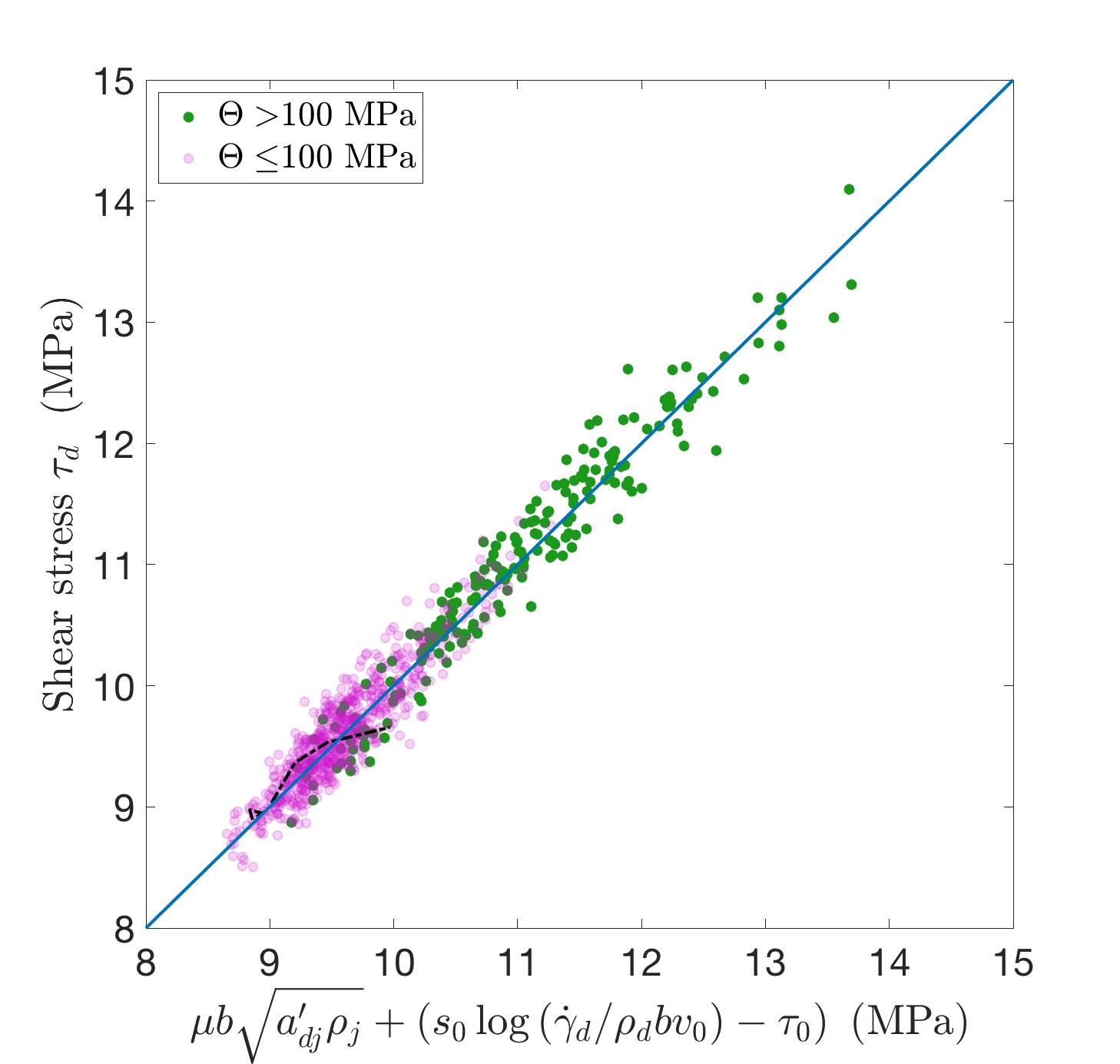}
        \caption{}
        \label{fig:Tau_logVi_adjRhoj_Hamed}
    \end{subfigure}
\caption[justification=centering]{Shear stress on the dominant slip system $\tau_d$ as a function of (a) $\alpha\mu b\sqrt{\rho}$, where $\rho$ is the total dislocation density, with $\alpha=0.37$,  (b) $\mu b \sqrt{a_{dj}\rho_j}$, where $a_{ij}$ are given in table~\ref{table:aij}, \nohl{and} (c) predictions from Eq.~(\ref{eq:taui_logVi_aijRhoj}). Data points for two sample low-hardening simulations are connected together by dashed lines. This figures contains results of $\approx 170$ DDD simulations of two different initial configurations, at strain rate $\dot{\varepsilon} = 10^3\,{\rm s}^{-1}$. 
\nohl{(RMSE, $R^2$) values are (a) (0.54~MPa, 0.62), (b) (0.48~MPa, 0.70) (solid line) and (0.40~MPa, 0.80) (dashed line), and (c) (0.20~MPa, 0.95)}. 
}
\label{fig:TaylorLaw} 
\end{figure}

Figure \ref{fig:Tau_totalRho} shows the comparison between Eq.~(\ref{Eq:Taylor}) and the DDD data, where $\alpha = 0.37$ and $\tau^{\rm rd} = 2.72$~MPa are the constant values that best fit the data.
The RMSE between this fit and all of the DDD data is $\textrm{RMSE}=0.54$ MPa.
It can be observed that the agreement is reasonably good for high-hardening orientations but not satisfactory for low-hardening orientations.
In particular, in the low-hardening orientations, the total dislocation density increases with strain due to dislocation multiplication, but the shear stress does not increase.  As a result, the data points from low-hardening DDD simulations move mostly parallel to the $x$-axis with increasing strain in figure \ref{fig:Tau_totalRho} (black dashed lines), instead of following the solid diagonal line delineating the Taylor relation. The similar issue was also reported in \cite{madec2002dislocation}, where it was corrected by multiplying a correction \nohl{factor} of $\ln\left(\frac{1}{b\sqrt{\rho_{\rm f}}}\right)$ to $\mu b\sqrt{\rho}$\nohl{, where $\rho_{\rm f}$ is the forest density,} to account for the dependence of the line-tension on dislocation screening. Here, including this term and obtaining the best fit to $\tau_d$, did not result in any improvement as the new RMSE was unchanged, still equal to 0.54 MPa.

Extensions to the classical Taylor law have been suggested to better describe the  strength of single crystals.
For example, it has been suggested that the forest dislocation density should be used instead of the total dislocation density~\cite{madec2002dislocation}.
Franciosi and Zaoui~\cite{franciosi1982multislip} suggested the following slip system-based extension of the Taylor relation,
\begin{equation}
  \tau_d = \mu b \sqrt{{a_{dj}\rho_j}} + \tau^{\rm rd}
  \label{eq:extended-Taylor}
\end{equation}
in which we have used Einstein's notation where the repeated index $j=1,...,12$ is summed over, and we have added the rate-dependent correction term $\tau^{\rm rd}$ as in Eq.~\eqref{Eq:Taylor}.
The dimensionless matrix coefficients $a_{ij}$ represent the strength of the interaction between slip systems $i$ and $j$, which, due to symmetry, have only six independent coefficients in FCC metals. 
\nohl{They} correspond to the four types of junction reactions between slip systems (Hirth, Lomer, glissile, colinear), and the coplanar and self interactions \cite{franciosi1982multislip}.
The interaction coefficients $a_{ij}$ were calculated in~\cite{madec2003role, devincre2006physical, kubin2008modeling} using specialized DDD simulations in which a dislocation in slip system $i$ glides through a preconstructed forest comprised of dislocations from slip systems $j$ that lead to a unique type of \nohl{interaction} with the system $i$.

\begin{table}
\centering
\small
\caption{Interaction coefficients between slip systems $i$ and $j$.  $a_{ij}$ from \cite{kubin2008modeling} and $a'_{ij}$ in this work by fitting Eq.~(\ref{eq:taui_logVi_aijRhoj}) to the DDD data.}
\label{table:aij}
\begin{tabular}{ P{4cm}  P{1.2cm}  P{1.2cm}  P{1.2cm}  P{1.2cm}  P{1.2cm}  P{1.2cm}    }
\toprule
  & $a_{\rm self}$ & $a_{\rm coplanar}$ & $a_{\rm Hirth}$ & $a_{\rm Lomer}$ & $a_{\rm Glissile}$ & $a_{\rm Collinear}$  \\
 \toprule
 \multirow{2}{4cm}{\centering $a_{ij}$ (from \cite{kubin2008modeling})  } &
 \multirow{2}{1.2cm}{\centering 0.122} &
 \multirow{2}{1.2cm}{\centering 0.122} &
 \multirow{2}{1.2cm}{\centering 0.07 }  &
 \multirow{2}{1.2cm}{\centering 0.122 } &
 \multirow{2}{1.2cm}{\centering 0.137 } &
 \multirow{2}{1.2cm}{\centering  0.625 } \\ \\ 
 \midrule
 \multirow{2}{4cm}{\centering  $a'_{ij}$ (this work)} & \multirow{2}{1.2cm}{\centering 0.300} & \multirow{2}{1.2cm}{\centering 0.152} & \multirow{2}{1.2cm}{\centering 0.083} & \multirow{2}{1.2cm}{\centering 0.326} & \multirow{2}{1.2cm}{ \centering 0.661} & \multirow{2}{1.2cm}{\centering 0.578} \\ \\
 \bottomrule
\end{tabular}
\end{table}

In figure~\ref{fig:Tau_aijRhoj_Kubin} we use the interaction coefficients $a_{ij}$ computed in~\cite{kubin2008modeling}, whose values are listed in table~\ref{table:aij}.
Here the rate-dependent term $\tau^{\rm rd}$ is treated again as a fitting constant.
The RMSE of the fit to Eq.~\eqref{eq:extended-Taylor} (\nohl{solid} line) is reduced to $\textrm{RMSE}=0.48$ MPa.
It can be seen that distribution of data points in figure~\ref{fig:Tau_aijRhoj_Kubin} suggests a slightly higher slope than $\mu b \sqrt{a_{ij}\rho_j}$ to obtain the best fit. Hence, a prefactor was also used to obtain the best linear fit as $\tau_d=1.20\mu b \sqrt{a_{dj}\rho_j}+2.42$, whose RMSE equals 0.40 MPa. This improvement when a prefactor is used, motivated us to obtain a new set of interaction coefficients, $a'_{ij}$, as will be detailed in the following paragraph.
For now we observe that the extended Taylor relation, Eq.~(\ref{eq:extended-Taylor}), shows better agreement with the DDD data than the classical Taylor relation as indicated by the lower RMSE.
In particular, Eq.~\eqref{eq:extended-Taylor} yields a \nohl{somewhat smaller discrepancy} for the low-hardening orientations because for these orientations, the increase of $a_{dj}\rho_j$ with strain is less rapid than the increase of the total dislocation density $\alpha^2\rho$.
Nonetheless, the discrepancy between Eq.~(\ref{eq:extended-Taylor}) (even \nohl{when} a prefactor is used) and the DDD data remains relatively large for low-hardening orientations.

By analyzing our data, we find that the extended Taylor relation can be significantly improved by replacing the rate-dependent constant $\tau^{\rm rd}$ with a term that explicitly depends on the plastic shear strain rate $\dot{\gamma}_d$ and dislocation density $\rho_d$ on the dominant slip system:
\begin{equation}
     \tau_d = \mu b \sqrt{a'_{dj}\rho_j}
             + \left[s_0 \log\left(\frac{\dot{\gamma}_d}{\rho_d b v_0}\right)-\tau_0 \right]
\label{eq:taui_logVi_aijRhoj}             
\end{equation}
where $s_0$, $v_0$ and $\tau_0$ are fitting coefficients (given in table~\ref{table:v0_s1}), and $a'_{ij}$ are interaction coefficients having the same symmetry but different numerical values than the $a_{ij}$ used previously. \nohl{We will justify the specific form we have chosen for Eq.~(}\ref{eq:taui_logVi_aijRhoj}\nohl{) in the next Section.}  The values for $a'_{ij}$ are given in table~\ref{table:aij}; here they are obtained by fitting Eq.~(\ref{eq:taui_logVi_aijRhoj}) to the data from strain hardening DDD simulations instead of using specialized simulations as in \cite{kubin2008modeling}. A similar approach was recently used to extract interaction coefficients in hcp Mg from DDD simulations \cite{messner2017crystal}. 
Overall, the relative ranking of the interaction coefficients $a_{ij}$ calculated in \cite{kubin2008modeling} and the fitted values $a'_{ij}$ obtained here are consistent, with the exception that for $a'_{ij}$ the glissile interaction has the largest value instead of the collinear interaction reported earlier \cite{madec2003role, kubin2008modeling}.
This result is consistent with recent work in which the glissile reaction was found to have the most important contribution to strain hardening \cite{sills2018dislocation}.

It can be seen from figure~\ref{fig:Tau_logVi_adjRhoj_Hamed} that the flow stress expression Eq.~(\ref{eq:taui_logVi_aijRhoj}) leads to a better agreement with the DDD data than Eq.~(\ref{eq:extended-Taylor}), especially for low-hardening orientations.
The \nohl{average error} in this case is $\textrm{RMSE}=0.20$ MPa, which is significantly reduced compared to the corresponding value in case of Eq.~(\ref{eq:extended-Taylor}).
\nohl{In comparison, if the second term in} Eq.~(\ref{eq:taui_logVi_aijRhoj}) is replaced by a constant, $\tau^{\rm rd}$, while leaving $a_{ij}'$ as free fitting parameters, then the resulting average error is RMSE = 0.32~MPa.  
\nohl{On the other hand, if the first term in} Eq.~(\ref{eq:taui_logVi_aijRhoj}) is constrained to be $\mu b \sqrt{a_{dj}\rho_j}$, while leaving $s_0$, $v_0$ and $\tau_0$ as free fitting parameters, then the resulting error is RMSE = 0.38~MPa.
\nohl{This shows that the modifications of both terms from} Eq.~(\ref{eq:extended-Taylor}) to Eq.~(\ref{eq:taui_logVi_aijRhoj}) are necessary for accurately describing the DDD data.

The physical interpretation of Eq.~(\ref{eq:taui_logVi_aijRhoj}) will be discussed in Section~\ref{sec:exponentialEqVelocity}.
Here we merely wish to point out that in Eq.~(\ref{eq:taui_logVi_aijRhoj}) the dislocation density appears in two distinct terms, representing two different effects.  In the first term, it represents a rate-independent strengthening mechanism (increasing dislocation density leads to higher stress), in which (forest) dislocations act as obstacles for the motion of other dislocations.
In the second term, it represents a rate-dependent softening mechanism (increasing dislocation density leads to lower stress), where (mobile) dislocations act as carriers of the plastic deformation.

As discussed in Section~\ref{sec:intro}, a rate-dependent flow stress expression also implies a flow rule.
Rewriting Eq.~(\ref{eq:taui_logVi_aijRhoj}) while assuming it applies to all slip systems $i$ gives an expression for the plastic flow rate $\dot{\gamma}_i$ on slip system $i$ of
\begin{equation}
     \frac{\dot{\gamma}_i}{\rho_i b} = v_0\exp \left[ \frac{1}{s_0}\left(\tau_i - \left(\mu b \sqrt{a'_{ij}\rho_j}-\tau_0\right) \right) \right]. 
\label{eq:vEff_expTaueff2}             
\end{equation}
\nohl{We shall see below that} Eq.~(\ref{eq:vEff_expTaueff2}) \nohl{is also in very good agreement with the DDD data.}
Eqs.~(\ref{eq:taui_logVi_aijRhoj}) and (\ref{eq:vEff_expTaueff2}) are \nohl{two of} the major results of this work. Given the Orowan relation, $\dot{\gamma}_i = \rho_i b \bar{v}_i$, where $\bar{v}_i$ is the average dislocation velocity on slip system $i$, the left hand side of Eq.~(\ref{eq:vEff_expTaueff2}) is simply $\bar{v}_i$.
Therefore, Eqs.~(\ref{eq:taui_logVi_aijRhoj}) and (\ref{eq:vEff_expTaueff2}) contains both the Taylor law (for strength) and the Orowan relation (for plastic flow).
The physical justification of Eqs.~(\ref{eq:taui_logVi_aijRhoj}) and (\ref{eq:vEff_expTaueff2}) provided by the DDD data will be presented in Sections~\ref{sec:Taylor-link} and \ref{sec:exponentialEqVelocity}.

\subsection{Origin of logarithmic rate dependence of the flow stress}
\label{sec:Taylor-link}

In this section we discuss how we arrived upon our final form of the flow stress Eq.~(\ref{eq:taui_logVi_aijRhoj}), and hence the flow rule in Eq.~(\ref{eq:vEff_expTaueff2}), and we propose a physical interpretation of our findings in relation to the microstructure. 
The logarithmic form of the rate dependent term in the flow stress is not trivial or obvious (to the authors, at least), \nohl{especially because our DDD simulations do not involve any thermally activated processes}.
The logarithmic dependence was initially discovered by attempting to identify correlations within the DDD database, e.g., by plotting $\tau_d$ with respect to various other quantities.
For example, Eq.~(\ref{eq:taui_logVi_aijRhoj}) indicates that $\tau_d$ is correlated with $\sqrt{a'_{dj}\rho_j}$ and with $\log{(\dot{\gamma}_d/\rho_d)}$.
\nohl{However, it would not have been straightforward to directly identify such a relation by plotting the DDD data, because $\sqrt{a'_{dj}\rho_j}$ depends on the values of the $a'_{ij}$ matrix, which are not known} {\it a priori}.
\nohl{Instead, we first discovered a correlation among $\tau_d$, $\log({\dot{\gamma}_d/\rho_d})$ and $1/\bar{l}_d$, where $\bar{l}_d$ is the average link length in the dominant slip system.}
Unlike $\sqrt{a'_{dj}\rho_j}$, $1/\bar{l}_d$ is straightforwardly extracted from the DDD database because it is strictly a geometric quantity characterizing the dislocation network.
Let us start this section by discussing the relationship between the \nohl{flow stress and the average link length} in the context of the Taylor law.
The Taylor relation is often rationalized in terms of the stress needed to move a dislocation link of length $l$---defined as a dislocation line pinned at both ends---following the Frank-Read (FR) source mechanism. Since the critical stress to activate such a  FR source is proportional to $\mu b / l$, and the average length of a dislocation link in a dislocation network of density $\rho$ is roughly proportional to $1/\sqrt{\rho}$, one arrives at the Taylor law, $\tau = \alpha \mu b \sqrt{\rho}$~\cite{saada1960hardening}.
In our DDD simulations however, dislocations do not multiply through the classical FR mechanism, \nohl{which was also pointed out in}~\cite{stricker2018dislocation}. 
Nonetheless, it is still of interest to examine if there exists a clear correlation between the resolved shear stress $\tau_d$ and the average link length $\bar{l}_d$ on the dominant slip system.

\begin{figure}[ht]
    \begin{subfigure}[t]{0.49\textwidth} 
    \includegraphics[scale=0.18,center]{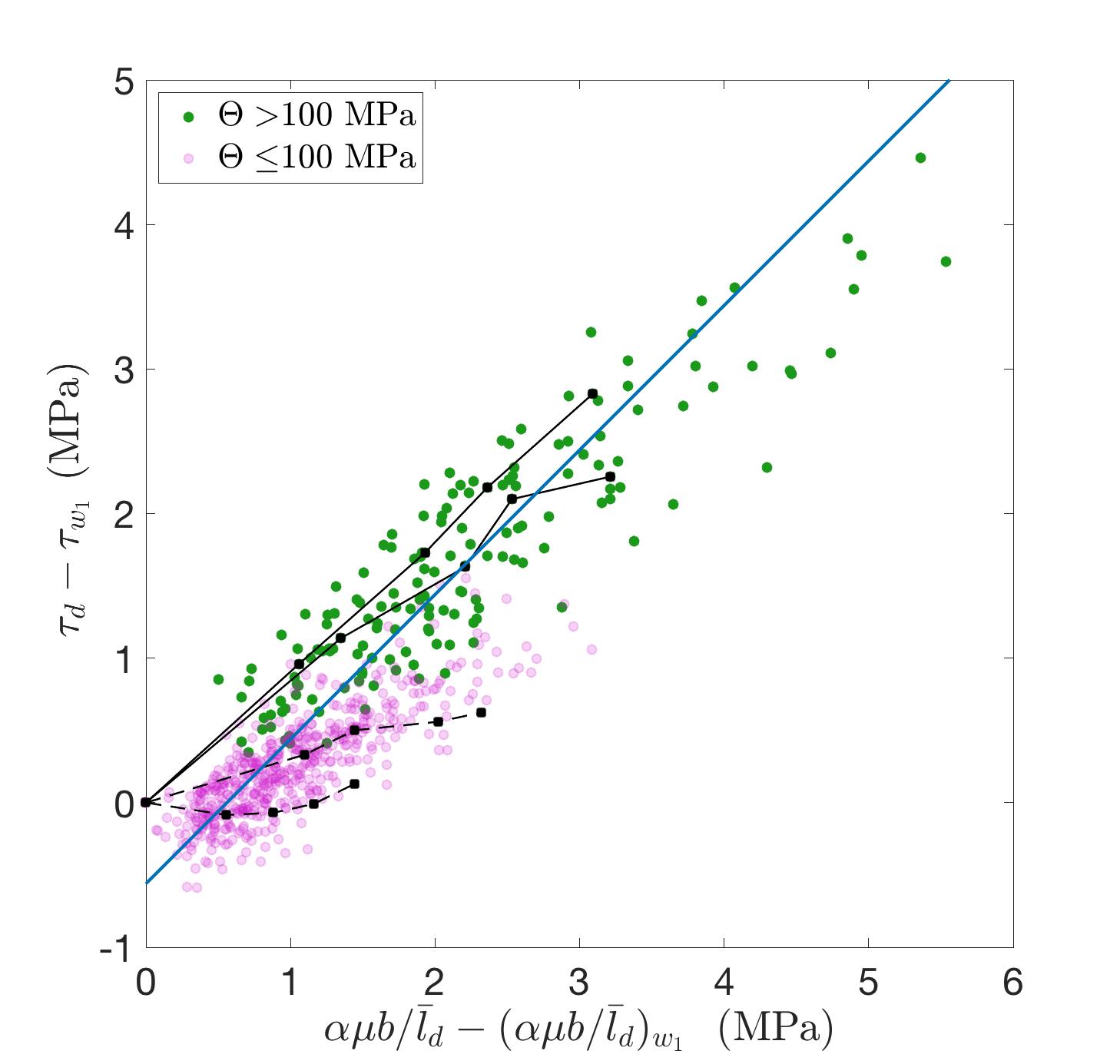}
    \caption{}
    \label{fig:taud_one_over_ld_A}
    \end{subfigure}
    \begin{subfigure}[t]{0.49\textwidth} 
    \includegraphics[scale=0.18,center]{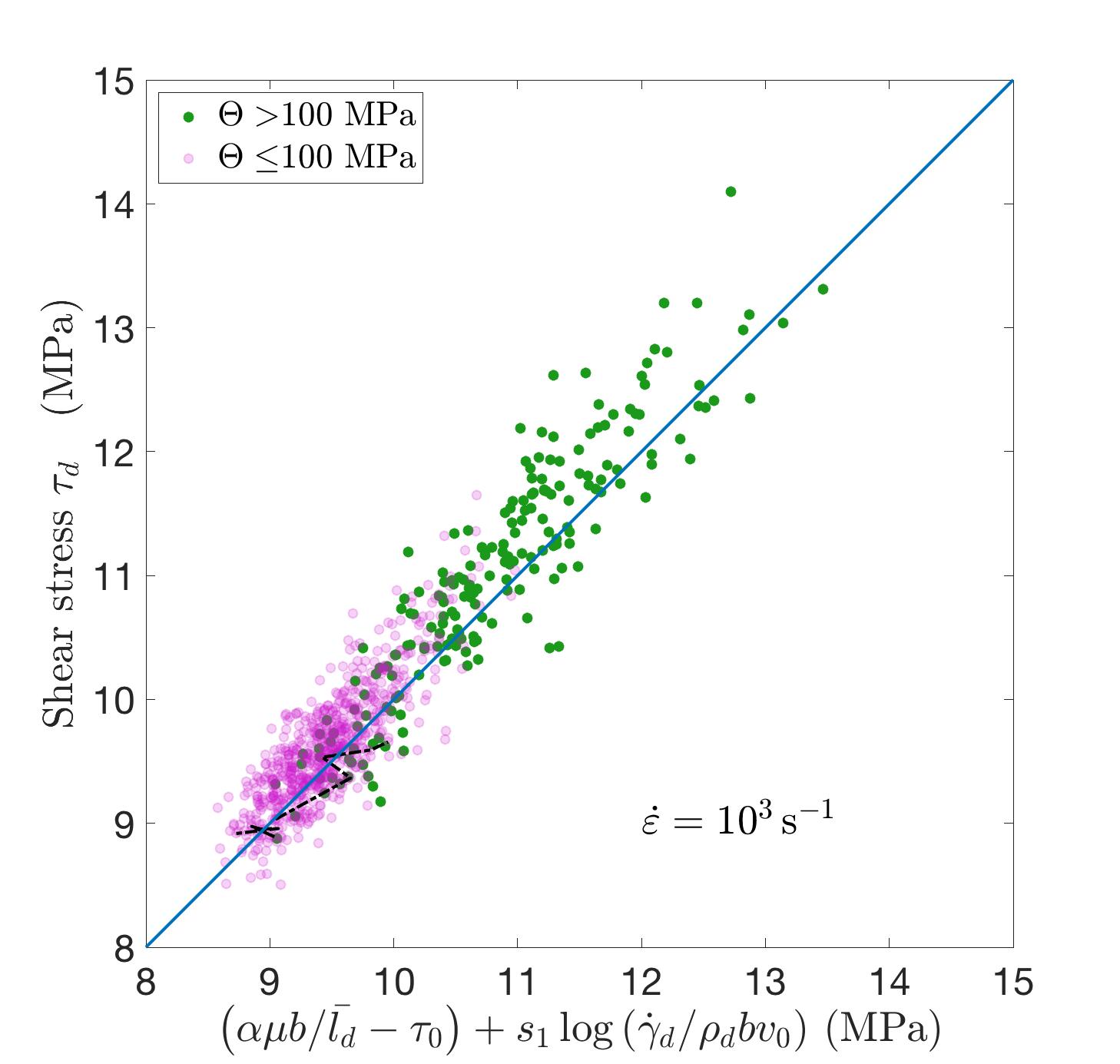}
    \caption{}
    \label{fig:taud_one_over_ld_B}
    \end{subfigure}
    \caption{(a) Correlation between the flow stress $\tau_d$ and ${\alpha \mu b}/{\bar{l}_d}$  during deformation. Data points were shifted by the value of the first block, denoted by $w_1$. \nohl{The solid line is the best fit to the data by a straight line $y=x-y_0$, where $y_0$ is a fitting constant, with error measures (RMSE, $R^2$) = (0.44~MPa, 0.66).}
    (b) $\tau_d$ predicted by DDD versus the corresponding prediction by Eq.~(\ref{eq:taud_mu_b_over_ld_logVd}) \nohl{with (RMSE, $R^2$) = (0.34~MPa, 0.85).}
    This figures contains results of $\approx 170$ DDD simulations of two different initial configurations,  at strain rate $\dot{\varepsilon} = 10^3\,{\rm s}^{-1}$. 
    }
    \label{fig:taud_one_over_ld}
\end{figure}

Figure~\ref{fig:taud_one_over_ld}a shows that the increase of $\tau_d$ during each DDD simulation (i.e., the strain hardening effect) can indeed be approximately described by the change in $\alpha \mu b/\bar{l}_d$, with $\alpha = 0.39$, especially for the high-hardening orientations.
The level of agreement is characterized by an average error of RMSE = \nohl{0.44~MPa}, which
is comparable to that of the extended Taylor relation Eq.~(\ref{eq:extended-Taylor}), as shown in figure~\ref{fig:Tau_aijRhoj_Kubin}.
In other words, the DDD data suggests the following relationship:
\begin{equation}
    \tau_d = \alpha \, \frac{\mu b}{\bar{l}_d} + \tau^{\rm rd}.
    \label{eq:taud_alpha_mu_b_over_ld}
\end{equation}
However, Eq.~(\ref{eq:taud_alpha_mu_b_over_ld}) is a stronger relation because it requires a single fitting parameter ($\alpha = 0.39$), compared to the six independent components of the $a_{ij}$ matrix.
Similar to our discussions of the Taylor law in Section~\ref{sec:modfTaylor}, we find that the agreement is further improved if the rate-dependent term is \nohl{expressed as a logarithmic function of the strain rate divided by the dislocation density} on the dominant system. 
This leads us to the following form of a modified Taylor law:
\begin{equation}
 \tau_d = \alpha \, \frac{\mu b}{\bar{l}_d}
        + \left[s_1 \log\left(\frac{\dot{\gamma}_d}{\rho_d b v_0}\right)-\tau_0 \right] 
\label{eq:taud_mu_b_over_ld_logVd}             
\end{equation}
where $s_1$, $v_0$, $\tau_0$ are fitting parameters (as given in table~\ref{table:v0_s1}).
\nohl{This expression was motivated by the almost linear relationship among $\tau_d$, $1/\bar{l}_d$, and $\log(\dot{\gamma}_d/\rho_d)$ values observed from our DDD data.}
Figure~\ref{fig:taud_one_over_ld}b shows that Eq.~(\ref{eq:taud_mu_b_over_ld_logVd}) indeed agrees very well with the DDD data.
Although its form is similar to Eq.~(\ref{eq:taui_logVi_aijRhoj}), Eq.~(\ref{eq:taud_mu_b_over_ld_logVd}) is a much stronger condition, because it only involves four fitting parameters ($\alpha$, $s_1$, $v_0$, $\tau_0$), compared with 9 in Eq.~(\ref{eq:taui_logVi_aijRhoj}), and yet it can describe the flow stress predictions from DDD simulations for over 120 different orientations with an averaged error of $\textrm{RMSE}=0.34$ MPa.
Because Eq.~\eqref{eq:taud_mu_b_over_ld_logVd} uses a single coefficient $\alpha$ to parametrize slip systems interactions for arbitrary loading orientations, it is less likely to be affected by the issue of over-fitting that may otherwise arise from the higher flexibility of the $a_{ij}$ matrix and the limited amount of DDD data.

\subsection{Derivation of a flow rule}
\label{sec:exponentialEqVelocity}

\begin{figure}[ht]
    \begin{subfigure}[t]{0.49\textwidth} 
         \includegraphics[scale=0.18,center]{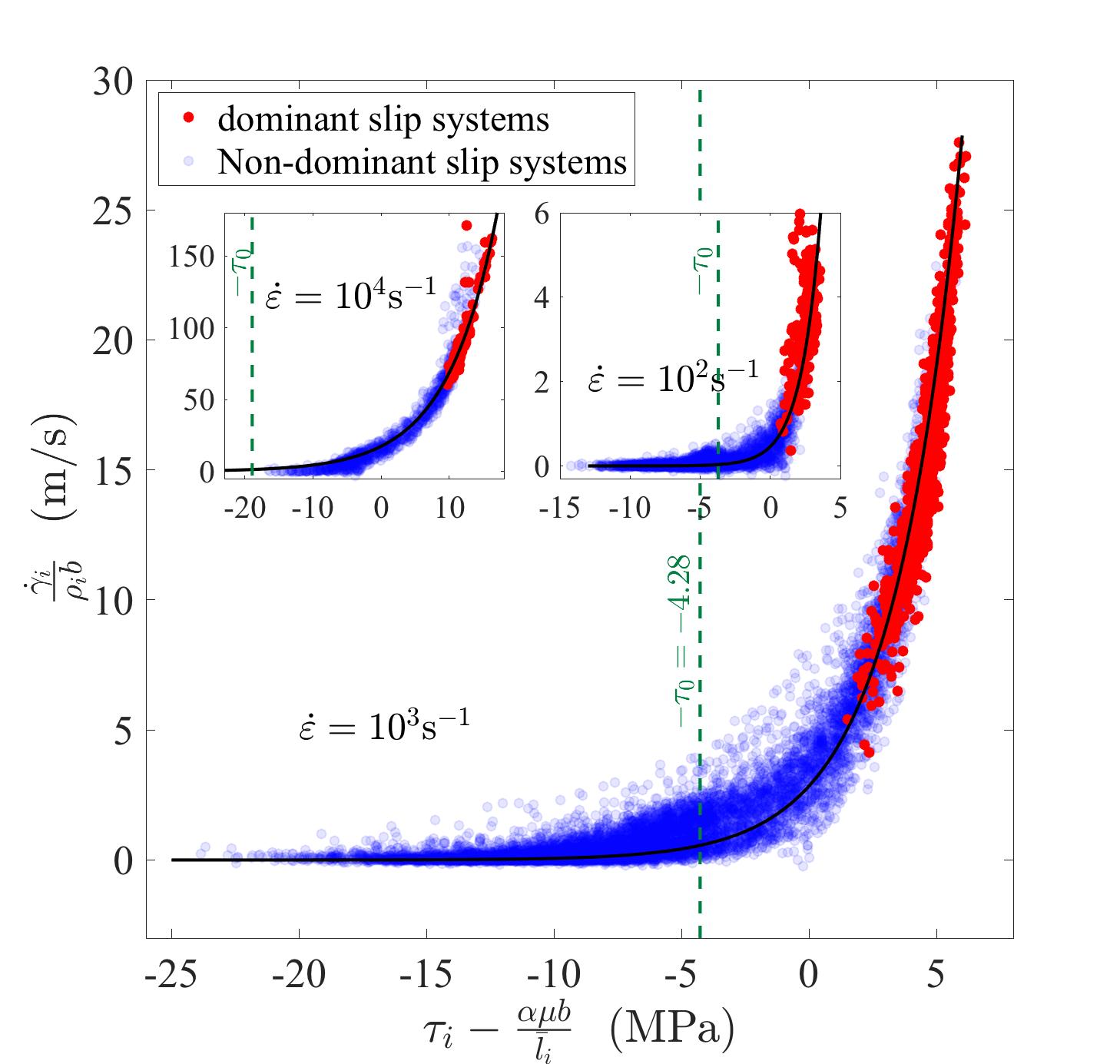}
        \caption{}
        \label{fig:Vel_TauEff}
    \end{subfigure}
    \begin{subfigure}[t]{0.49\textwidth} 
        \includegraphics[scale=0.18,center]{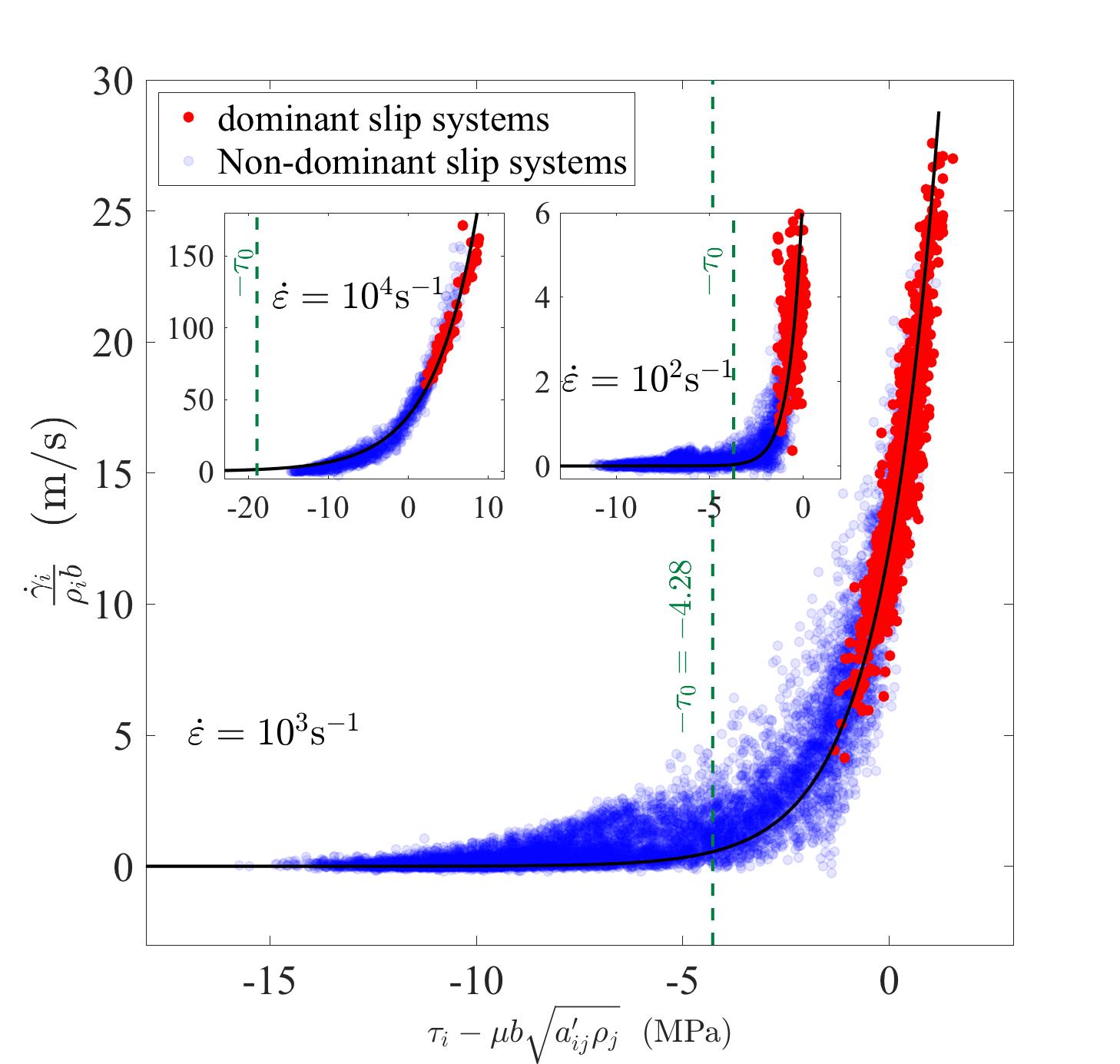}
        \caption{}
        \label{fig:Vel_tauEff2}
    \end{subfigure}
    \caption{(a) Exponential relationship between ${\dot{\gamma}_i}/({\rho_i b})$ and $\tau_i - \alpha\mu b/\bar{l}_i$, on every slip system $i$,  as stated in Eq.~(\ref{eq:vEff_expTaueff}), in solid line.
    \nohl{The error between fitted curve and DDD data is (RMSE, $R^2$) =  (0.37~$\rm{m/s}$, 0.87),  (1.09~$\rm{m/s}$, 0.95), and  (8.96~$\rm{m/s}$, 0.94)
    for $\dot{\varepsilon} = 10^2\,\rm{s}^{-1}, 10^3\,\rm{s}^{-1}$ and $10^4\,\rm{s}^{-1}$, respectively.}
    (b) Exponential relationship between ${\dot{\gamma}_i}/({\rho_i b})$ and $\tau_i-\mu b \sqrt{a'_{ij}\rho_j}$, on every slip system $i$,  as stated in Eq.~(\ref{eq:vEff_expTaueff2}), in solid line. 
    \nohl{The error of fitting (RMSE, $R^2$) = (0.52~$\rm{m/s}$, 0.74),  (1.40~$\rm{m/s}$, 0.92), and  (7.66~$\rm{m/s}$, 0.96) for $\dot{\varepsilon} = 10^2\,\rm{s}^{-1}, 10^3\,\rm{s}^{-1}$ and $10^4\,\rm{s}^{-1}$, respectively.}
    Dots are data points correspond to total of $\approx$ 170 DDD simulations, from two different initial configurations, under strain rate $\dot{\varepsilon}=10^3\,{\rm s^{-1}}$.
    The insets contain data points from 54 and 18 DDD simulations under strain rate $\dot{\varepsilon}=10^2\,{\rm s^{-1}}$ and $10^4\,{\rm s^{-1}}$, respectively. 
    }
    \label{fig:exponential_flow_rule}
\end{figure}

As discussed in Section~\ref{sec:intro}, the flow rule implied by any rate-dependent flow stress expression can be obtained by solving for the shear strain rate.
Performing such a manipulation of Eq.~(\ref{eq:taud_mu_b_over_ld_logVd}), and while generalizing it to all slip systems $i$ gives
\begin{equation}
     \frac{\dot{\gamma}_i}{\rho_i b} 
     = v_0 \exp{\left[ \frac{1}{s_1}
        \left(\tau_i-\left(\frac{\alpha \mu b}{\bar{l}_i}-\tau_0\right) \right) \right] }.
      \label{eq:vEff_expTaueff}
\end{equation}
Similar to Eqs.~(\ref{eq:taui_logVi_aijRhoj}) and (\ref{eq:vEff_expTaueff2}), 
Eqs.~(\ref{eq:taud_mu_b_over_ld_logVd}) and (\ref{eq:vEff_expTaueff}) combine both the Taylor law and the Orowan relation in one expression.
It can be seen from figure~\ref{fig:Vel_TauEff} that Eq.~(\ref{eq:vEff_expTaueff}) is in very good agreement with the DDD data on all slip systems over all loading orientations, with \nohl{RMSE = 1.09 ${\rm m}\cdot{\rm s}^{-1}$}.
\nohl{A qualitative argument for the exponential dependence of average dislocation velocity $\bar{v}_i$ on $\tau_i - \alpha \mu b/\bar{l}_i$ can be given as follows.}
\nohl{First, $\alpha \mu b/\bar{l}_i$ can be regarded as a ``critical stress'' to initiate plastic flow on slip system $i$, if all dislocation links have the same length $\bar{l}_i$.}
However, all of the dislocation links on slip system $i$ do not have the same length (\nohl{leading to $\tau_0$ in} Eq.~(\ref{eq:vEff_expTaueff})).
Instead, from~\cite{sills2018dislocation} we may expect that the dislocation link length satisfies an exponential distribution, with the probability of finding a dislocation link with length much longer than the average length being exponentially small.
At very small stress $\tau_i$, only a very small fraction of dislocation links (whose lengths are greater than $\alpha\mu b/\tau_i$) are activated and contribute to plastic flow.
\nohl{With increasing $\tau_i$, the fraction of dislocation links that become activated increases exponentially fast, leading to an exponential dependence of $\dot{\gamma}_i$ on $\tau_i$.}
\nohl{However, we would like to point out that the ultimate justification of} Eq.~(\ref{eq:taud_mu_b_over_ld_logVd}) \nohl{comes from its agreement with the DDD data; we are currently unable to give a concise derivation of} Eq.~(\ref{eq:taud_mu_b_over_ld_logVd}) \nohl{using a simple model based on the idea outlined above---this is the subject of future work.}

\begin{table} 
\centering
\small
\caption{Fitting coefficients in the proposed flow rule shown in Eqs.~(\ref{eq:taud_mu_b_over_ld_logVd}) and (\ref{eq:vEff_expTaueff}).
} \label{table:v0_s1}
\begin{tabular}{ P{3cm}  P{1.6cm}  P{1.2cm} P{1.2cm} P{1.2cm}  P{1.2cm} }
\toprule
 & $v_0\,\rm(m\cdot s^{-1})$ & $s_0\,(\rm MPa)$ & $s_1\,(\rm MPa)$ & $\alpha$ & $\tau_0\,(\rm MPa)$ \\
\toprule
$\dot{\varepsilon}=10^2$\,s$^{-1}$ & 0.033 & 0.70 & 1.40 & 0.39 & 3.71 \\
\midrule
$\dot{\varepsilon}=10^3$\,s$^{-1}$ & 0.557 & 1.39 & 2.62 & 0.39 & $4.28$ \\
\midrule
$\dot{\varepsilon}=10^4$\,s$^{-1}$ & 0.775 & 5.60 & 7.30 & 0.39 & 18.91  \\
\bottomrule
\end{tabular}
\end{table}

The exponential form of the flow rule proposed in Eq.~(\ref{eq:vEff_expTaueff}) should be contrasted with other forms commonly used in CP modeling. 
One common example is the power-law flow rule, Eq.~(\ref{eq:powerlaw}), \nohl{where in the present context $\tau_{i}^{\rm c} = \alpha\mu b/\bar{l}_i$}. 
While the power-law flow rule implies a multiplicative decomposition of the rate-independent and rate-dependent contributions of the flow stress, Eq.~(\ref{eq:invpowerlaw}), the flow rule obtained here is instead derived from an additive decomposition according to Eq.~(\ref{eq:taud_mu_b_over_ld_logVd}).
See Appendix~\ref{appx:powerLaw} \nohl{for more discussions on why the multiplicative decomposition of the flow stress is not consistent with our DDD data.}
In the literature, \nohl{similar additive flow rules have also been proposed} \cite{franccois2012mechanical,steinberg1988constitutive}, \nohl{as well as}
exponential relationships between $\dot{\gamma}_d$ and $\tau_d$ which are based on thermally activated dislocation motion~\cite{frost1971motion,kocks1975thermodynamics, busso1996dislocation,ma2004constitutive}. 
However, we note that the exponential form in Eq.~(\ref{eq:vEff_expTaueff}) does not find its physical origin in thermal activation, because a simple, linear mobility law is used in our DDD simulations.
Furthermore, the over-damped equations of motion used in DDD \cite{arsenlis2007enabling} prohibit thermal fluctuations, thereby suppressing the possibility of thermal activation in governing the dynamics of the dislocation network.
Instead, the exponential form of the flow rule seen here must be the result of interactions within the dislocation network.

The insets in figure~\ref{fig:Vel_TauEff} show that Eq.~(\ref{eq:vEff_expTaueff}) is still in good agreement with the DDD data under uniaxial applied strain rates of $\dot{\varepsilon} = 10^2 \, \rm s^{-1}$ and $10^4 \, \rm s^{-1}$, provided that different fitting coefficients are used for $v_0$, $s_1$ and $\tau_0$ (as given in table~\ref{table:v0_s1}).
It is worth noting that the same value for $\alpha$ can be used for all three strain rates, indicating that the rate-independent contribution of the flow stress (inspired by the Taylor relation) is well accounted for in Eq.~(\ref{eq:taud_alpha_mu_b_over_ld}).
Table~\ref{table:v0_s1} shows that the magnitude of $s_1$ decreases with decreasing applied strain rate $\dot{\varepsilon}$.  
\nohl{This means that as the overall applied strain rate is decreased, the plastic strain rate on slip systems $i$ becomes more sensitive to the ``effective stress'', $\tau_i-\alpha\mu b/\bar{l}_i$.} 
This trend can be seen in the insets in figure \ref{fig:exponential_flow_rule}a.
Consequently, at lower applied strain rates, the flow stress becomes less sensitive to the plastic shear strain rate.
Therefore, in the quasi-static limit ($\dot{\varepsilon} \to 0$), we expect the flow stress of \nohl{pure} FCC metals \nohl{(in the absence of thermally activated processes)} to be independent of the plastic shear strain rate and thus solely governed by forest interactions, as expected from the classic Taylor law (i.e., $\tau^{\rm rd} \to 0$).
\nohl{In this limit, rate-dependent crystal plasticity should also be replaced by the classical, rate-independent plasticity formulation, in which the plastic strain rate is controlled by the boundary condition.}

The success of Eq.~(\ref{eq:vEff_expTaueff}) motivated us to propose a similar relationship linking $\dot{\gamma}_i$, $\tau_i$ and $\rho_i$.
Recall that our goal in this paper is to construct the simplest possible constitutive model for strain hardening that is consistent with the DDD data.
In doing so, we have limited our microstructural state variables to the dislocation densities, $\rho_i$, on the 12 slip systems.
If we were to account for both $\rho_i$ and $\bar{l}_i$ for each slip system, \nohl{as would be required to make use of} Eqs.~\eqref{eq:taud_mu_b_over_ld_logVd} and \eqref{eq:vEff_expTaueff}, the model would have 24 state variables, and would become much more complex.
In such a model, one would need to specify the evolution equation not only for $\dot{\rho}_i$ but also for $\dot{\bar{l}}_i$.  That is beyond the scope of this paper, but may be a possibility for future study.
Therefore, we propose the flow rule as given in Eq.~(\ref{eq:vEff_expTaueff2}), reproduced below,
\begin{equation*}
     \frac{\dot{\gamma}_i}{\rho_i b} = v_0\exp \left[ \frac{1}{s_0}\left(\tau_i - \left(\mu b \sqrt{a'_{ij}\rho_j}-\tau_0\right) \right) \right] 
\end{equation*}
where $v_0$, $s_0$ and $\tau_0$ are fitting constants (as given in Table~\ref{table:v0_s1}).  We note that for consistency $v_0$ and $\tau_0$ are the same as those in Eq.~(\ref{eq:vEff_expTaueff}); however the coefficient $s_0$ is different from the $s_1$ coefficient used in Eq.~(\ref{eq:vEff_expTaueff}).
It can be seen from figure~\ref{fig:Vel_tauEff2} that Eq.~(\ref{eq:vEff_expTaueff2}) is in good agreement with the DDD data for all loading orientations and three applied strain rates.
Moreover, figure~\ref{fig:exponential_flow_rule} also includes results from 54 DDD simulations using the second initial configuration, which confirm the validity and robustness of Eqs.~(\ref{eq:vEff_expTaueff}) and (\ref{eq:vEff_expTaueff2}). 
The best agreement is obtained for dominant slip systems (red dots).  On non-dominant slip systems (blue dots), in the regime of $\tau_i - \mu b\sqrt{a'_{ij}\rho_j} < -\tau_0$ where Eq.~(\ref{eq:vEff_expTaueff2}) predicts a vanishingly small shear strain rate, there are many cases where the shear strain rate predicted by the DDD simulations are non-negligible.
This discrepancy seems more significant in figure~\ref{fig:Vel_tauEff2} than in figure~\ref{fig:Vel_TauEff}.
Nonetheless, we think this is a reasonable compromise given the simplicity of the constitutive model that we are developing here.

\nohl{In summary, we have described the approach that led us to the unified expressions containing the logarithmic strain rate-dependence for the flow stress and the exponential stress-dependence for the flow rate, as given in} Eqs.~(\ref{eq:taui_logVi_aijRhoj}) and (\ref{eq:vEff_expTaueff2}).
\nohl{These relations are motivated by stronger relations,} Eqs.~(\ref{eq:taud_mu_b_over_ld_logVd}) and (\ref{eq:vEff_expTaueff}), \nohl{which involve fewer fitting parameters, and hence are expected to be more physically justifiable.} 
\nohl{Given that all of the findings of this work are based on DDD simulation data at strain rates above $10^2$\,${\rm s}^{-1}$, our conclusions are only supported in the high strain rate regime.}

\subsection{Kocks-Mecking model for dislocation multiplication} 
\label{sec:rhoidot}

Having found an expression for the shear strain rate $\dot{\gamma}_i$, our next step in developing a self-consistent constitutive model for strain hardening is to find an appropriate expression for the dislocation multiplication rate $\dot{\rho}_i$ that is consistent with the DDD data.
To do so, we start with the Kocks-Mecking model~\cite{mecking1981kinetics}, which expresses the rate of change of total dislocation density $\rho$ with shear strain $\gamma$ in the following form:
\begin{equation}
       \frac{d\rho}{d\gamma} = \frac{c_1}{b} \sqrt{\rho}-c_2\rho 
     \label{eq:KocksMeck_orig}
\end{equation}
where $c_1$ and $c_2$ are dimensionless constants.
The first term, characterizing dislocation multiplication, is taken to be inversely proportional to the mean free path of dislocations, $\Lambda \propto 1/\sqrt{\rho}$, defined as the distance travelled by a dislocation ``segment'' before it is stored~\cite{devincre2008dislocation}.
The second term, originally introduced to account for dynamic recovery~\cite{kocks2003physics}, represents dislocation annihilation and is assumed to be proportional to $\rho$~\cite{mecking1970new,essmann1979annihilation,mecking1981kinetics}.

\begin{figure}[ht]
    \begin{subfigure}[t]{0.33\textwidth} 
        \includegraphics[trim={0 0 0cm 0},clip=true,scale=0.135,right]{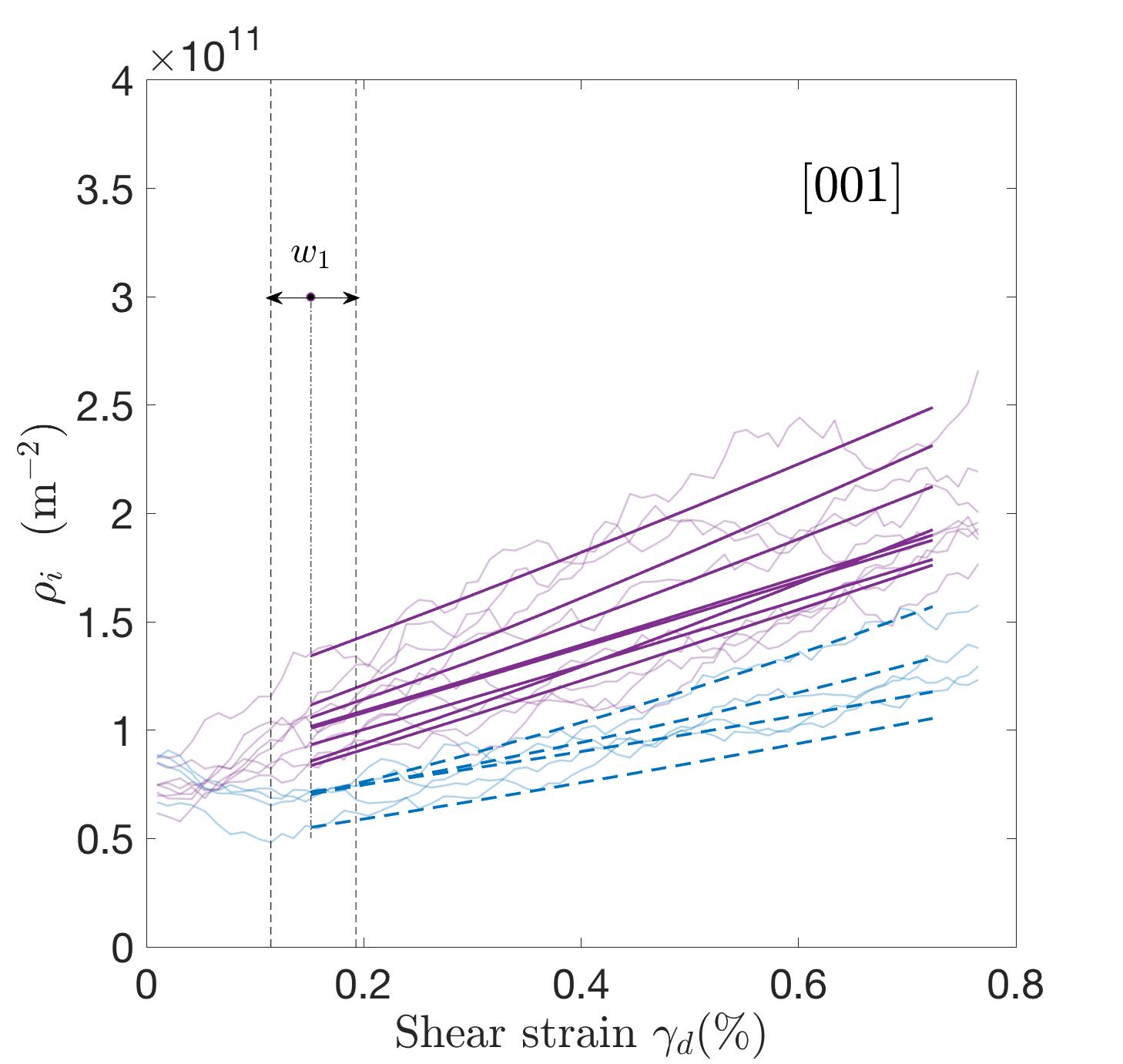}
        \caption{}
    \label{fig:Rhoi_001}
    \end{subfigure}\hfill
    \begin{subfigure}[t]{0.33\textwidth} 
        \includegraphics[trim={3.6cm 0 0 0},clip=true,scale=0.135,center]{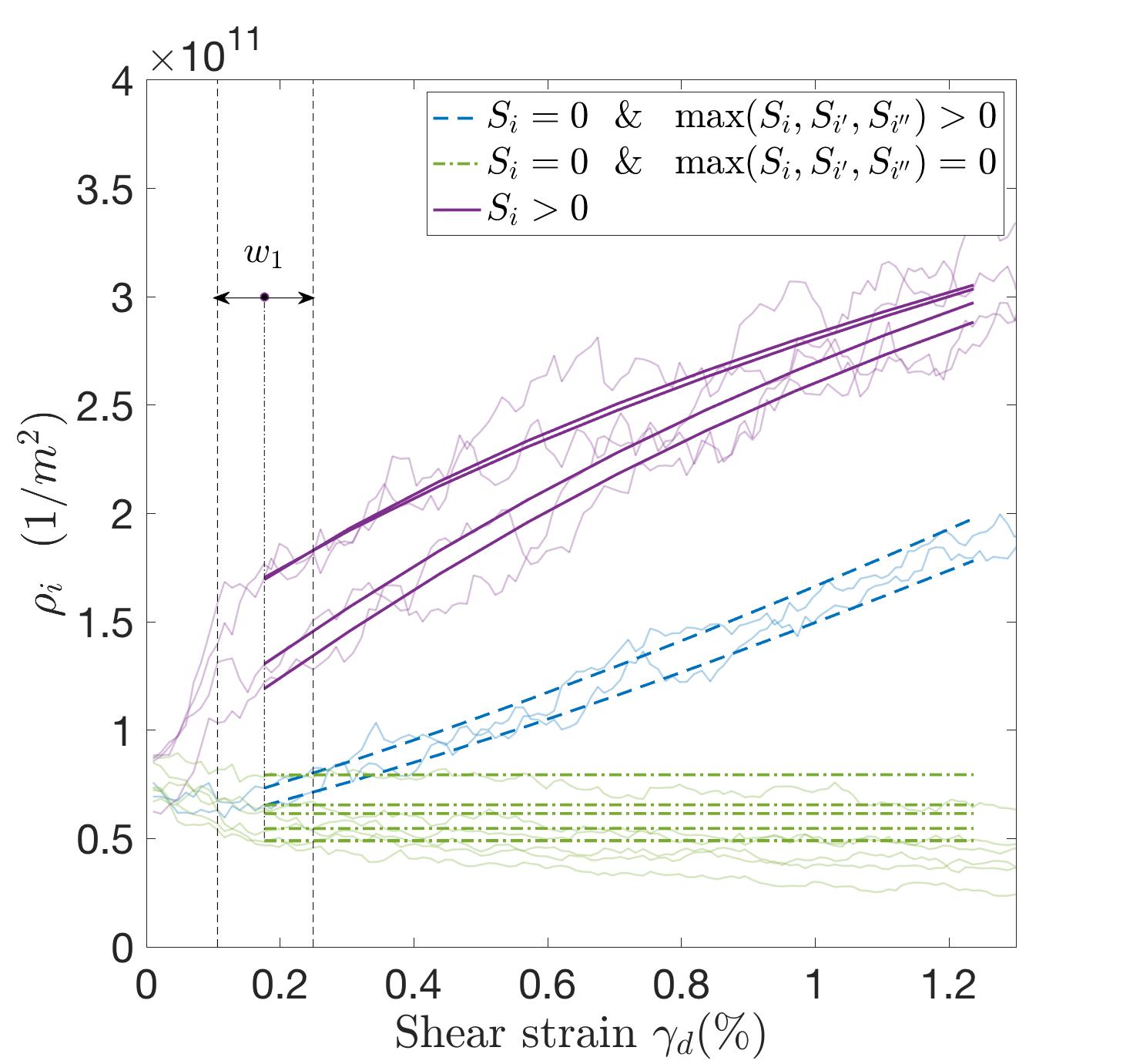}
        \caption{}
        \label{fig:Rhoi_011}
    \end{subfigure}
    \begin{subfigure}[t]{0.33\textwidth} 
        \includegraphics[trim={3.6cm 0 0 0},clip=true,scale=0.135,left]{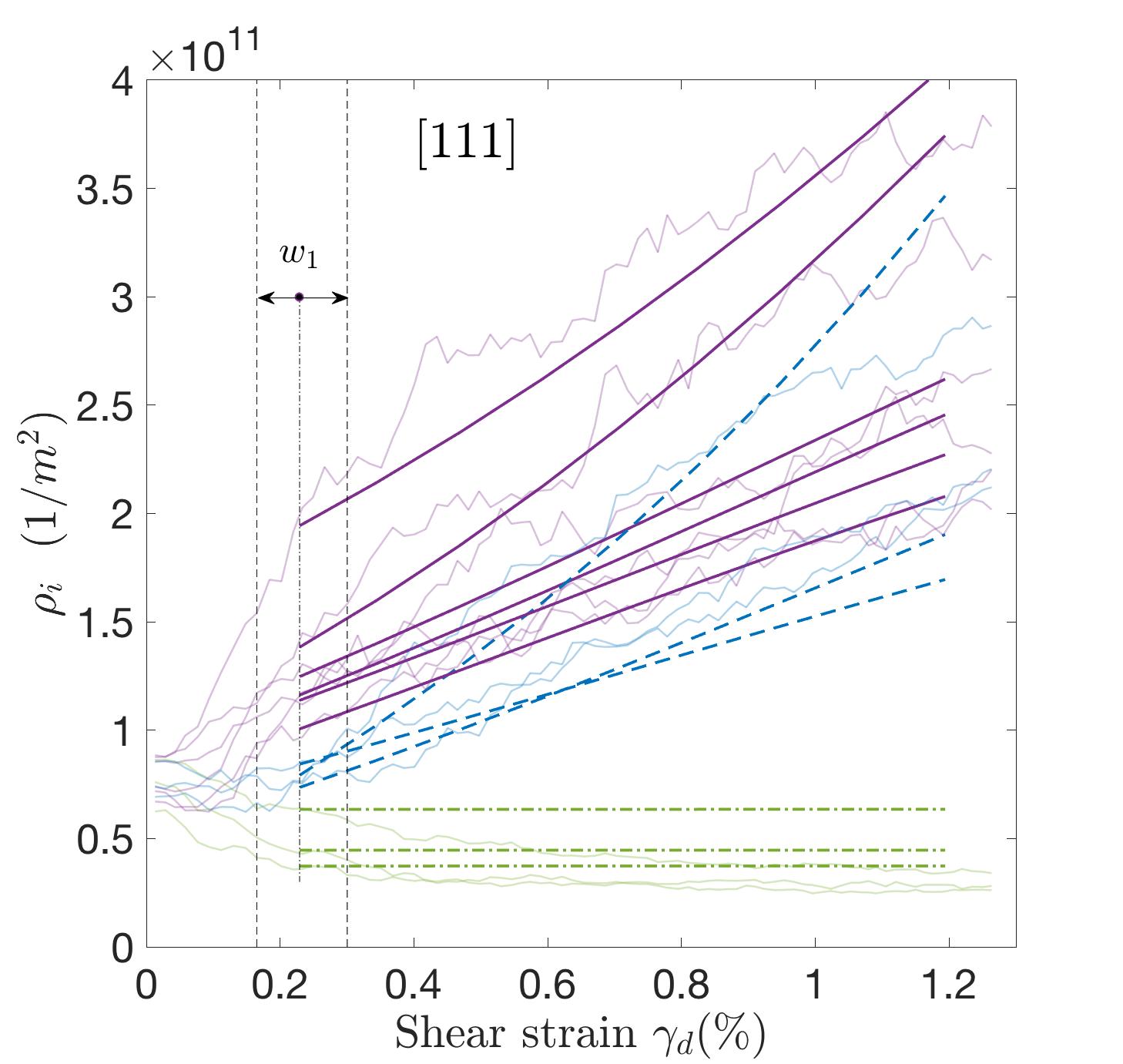}
        \caption{}
        \label{fig:Rhoi_111}
    \end{subfigure}
  \caption[justification=centering]{ Dislocation density on individual slip systems $\rho_i$ for DDD simulations along (a) $[001]$, (b) $[011]$ and (c) $[111]$ loading orientations. Simulation values are shown by thinner lines, while the best fit using Eqs.~(\ref{eq:KocksMeck_proposed}) and (\ref{eq:Rhoi_corr}) are shown by thicker lines. Values at the mid-point of the first block, ${w}_1$, were used as the initial values for integration. Active slip systems are shown by solid lines. Slip systems with zero Schmid factor that are coplanar with an active system are shown by dashed lines; the remaining slip systems with zero Schmid factor are shown by dot-dashed lines.  The fitting coefficients are (a) $(\tilde{c}_1, \tilde{c}_2, \tilde{c}_3)= (6.52\times 10^{-2}, 581, 2.91\times 10^{-2})$, (b) $(\tilde{c}_1, \tilde{c}_2, \tilde{c}_3)=(6.70\times 10^{-2}, 614, 1.22\times 10^{-2})$ and (c) $(\tilde{c}_1, \tilde{c}_2, \tilde{c}_3)=(3.16\times 10^{-2}, 213, 2.75\times 10^{-2})$. }
\label{fig:rhoi_gammad} 
\end{figure}

We follow Devincre et al.~\cite{devincre2008dislocation} to generalize the original Kocks-Mecking model to dislocation multiplication on individual slip systems with the following expression
\begin{equation}
     \dot{\rho}_i = \dot{\gamma}_i\left(\frac{\tilde{c}_1}{b}\sqrt{a'_{ij}\rho_j}-\tilde{c}_2\rho_i \right)+\dot{\rho}_{i}^{\rm corr}(\{\dot{\gamma}_j\}, \{\rho_j\})  
     \label{eq:KocksMeck_proposed}
\end{equation}
where $\tilde{c}_1$ and $\tilde{c}_2$ are dimensionless fitting coefficients that are constrained to be positive.  We added the $\tilde{}$ symbol to $\tilde{c}_1$ and $\tilde{c}_2$ to signify the fact that they will be dependent on the loading orientation (but still independent of the slip system index $i$).  This is in contrast to all the other fitting parameters in this work, such as $\alpha$, $v_0$, $s_0$, $a'_{ij}$, for which the same values are used for all loading orientations encountered in the DDD simulations.
We note that the mean free path coefficients of Devincre et al.~\cite{devincre2008dislocation} are also orientation-dependent.
\nohl{The loading-orientation dependence of $\tilde{c}_1$ and $\tilde{c}_2$ coefficients suggests that} Eq.~(\ref{eq:KocksMeck_proposed}) \nohl{still treats some important physics regarding dislocation multiplication in a phenomenological way.}
It is possible that this orientation dependence can be accounted for by introducing suitable functions of the Schmid factors.  But this possibility will be left to future studies. 
\nohl{We wish to point out that even for a specific loading orientation, being able to characterize the dislocation density evolution over 12 slip systems over the entire range of strain using a limited number of fitting coefficients (such as $\tilde{c}_1$ and $\tilde{c}_2$), is, we believe, not a trivial finding.  In this sense the modified Kocks-Mecking model given by Eq.~\eqref{eq:KocksMeck_proposed} is still useful, although more work is needed to further improve our understanding and modelling of dislocation multiplication.}
\nohl{For example, it is unclear how to generalize the model developed here for uniaxial loading to multiaxial loading, or loading where the load path changes in time (non-proportional loading).}
\nohl{The interaction coefficients, $a'_{ij}$, used in} Eq.~\eqref{eq:KocksMeck_proposed}, are the same as the ones in Eq.~\eqref{eq:taui_logVi_aijRhoj}. There are some examples in the literature where different interaction coefficients are used in the equations for the flow stress and dislocation multiplication, respectively~\cite{casals2007crystal}.

In Eq.~(\ref{eq:KocksMeck_proposed}), $\dot{\rho}_{i}^{\rm corr}$ is a correction term we added to the previous dislocation multiplication model~\cite{devincre2008dislocation}.
We found it necessary because if we were to strictly follow the classical Kocks-Mecking model, the dislocation multiplication rate $\dot{\rho}_i$ would be zero if its plastic strain rate $\dot{\gamma}_i$ were zero.
However, in our DDD simulations we have found many cases in which a slip system $i$ with zero Schmid factor $S_i$ and zero plastic shear strain rate $\dot{\gamma}_i$ nonetheless have an appreciable dislocation multiplication rate $\dot{\rho}_i$ \nohl{(as high as 50\% of the $\dot{\rho}_d$)}; \nohl{we refer to this as \emph{slip-free multiplication}}.
This was reported in previous DDD~\cite{weygand2014mechanics, stricker2015dislocation} and molecular dynamics simulations~\cite{zepedaruiz2019metal} (see Section~\ref{sec:glissilejunction} for more discussions).
In the discussion below, we denote the two slip systems which are coplanar to slip system $i$ as slip systems $i'$ and $i''$.
\nohl{Slip-free multiplication} is most pronounced along the three high symmetry loading orientations, $[001], [011]$ and $[111]$, as shown in figure~\ref{fig:rhoi_gammad}.
With all three of these orientations, we find that slip-free multiplication occurs if they are coplanar to an active slip system.
Under $[001]$ loading (figure~\ref{fig:rhoi_gammad}a) all four slip systems with $S_i=0$ are coplanar to slip systems with $S_{i'}=S_d=0.41$, so they all exhibit non-negligible multiplication.
With $[011]$ loading (figure~\ref{fig:rhoi_gammad}b), only two of the eight slip systems with $S_i=0$ are coplanar to systems with $S_{i'}=S_d=0.41$, and only these two slip systems among the eight exhibit non-negligible multiplication.
Finally, with $[111]$ loading (figure~\ref{fig:rhoi_gammad}c) the same behavior is observed, showing multiplication in three of the six slip systems with $S_i=0$ which are coplanar to slip systems with $S_i = S_d = 0.27$.

From these observations, it appears that slip-free multiplication is possible within a slip system if another slip system (or two slip systems) on the same plane has non-zero plastic strain rate.
Therefore, we propose the following correction term to the generalized Kocks-Mecking model for dislocation multiplication, Eq.~(\ref{eq:KocksMeck_proposed}):
\begin{equation}
    \dot{\rho}_{i}^{\rm corr}(\{\dot{\gamma}_j\}, \{\rho_j\}) 
    = \frac{\tilde{c}_3}{b} \left(\dot{\gamma}_{i'}\sqrt{\rho_{i''}}+ \dot{\gamma}_{i''}\sqrt{\rho_{i'}} \right)
    \label{eq:Rhoi_corr}
\end{equation}
where $i'$ and $i''$ are the two slip systems co-planar with slip system $i$ and $\tilde{c}_3$ is a dimensionless fitting parameter (that depends on loading orientation).

As shown in figure~\ref{fig:rhoi_gammad}, the fluctuations in $\rho_i$ are rather significant during the DDD simulations, making it difficult to directly extract $\dot{\rho}_i$.
In order to determine the fitting parameters $\tilde{c}_1$, $\tilde{c}_2$ and $\tilde{c}_3$, in Eqs.~(\ref{eq:KocksMeck_proposed}) and (\ref{eq:Rhoi_corr}), we obtain the predicted $\rho_i(t)$ curves by numerically integrating Eq.~(\ref{eq:KocksMeck_proposed}) using trial values of $\tilde{c}_1$, $\tilde{c}_2$ and $\tilde{c}_3$.
We shall use $\rho^{\rm nint}_i(t)$ to represent the numerically integrated density functions, which depend on parameters $\tilde{c}_1$, $\tilde{c}_2$ and $\tilde{c}_3$.
The initial values of $\rho_i$ for the numerical integration are the averaged values from DDD data in the first time block, \nohl{denoted as $w_1$ in} figure~\ref{fig:SigmaEps}b (see Section~\ref{sec:DDDsetup}).  In this section, the $\dot{\gamma}_i$ values needed in the numerical integration of Eqs.~(\ref{eq:KocksMeck_proposed}) and (\ref{eq:Rhoi_corr}) are also taken from the DDD data, as described in Section~\ref{sec:DDDsetup}.
The parameters $\tilde{c}_1$, $\tilde{c}_2$ and $\tilde{c}_3$ are determined by minimizing the mean square error between $\rho^{\rm nint}_i(t)$ and averaged values from the DDD data, i.e., by minimizing the following loss function
\begin{equation}
    L(\tilde{c}_1,\tilde{c}_2,\tilde{c}_3) = \sum_{n=1}^{9} \sum_{i=1}^{12}{ \left[ \rho_i^{\rm DDD}(t_n)-\rho_i^{\rm nint}(t_n; \tilde{c}_1,\tilde{c}_2,\tilde{c}_3) \right]^2}
    \label{eq:Loss_rhoi}
\end{equation}
\nohl{where $n$ is summed over the 9 averaging blocks for each DDD simulation data} (as explained in Section~\ref{sec:DDDsetup}), $t_n$ is the mid-time of the block $w_n$.
A gradient descent algorithm is implemented for the minimization of function $L$.
It can be seen from figure~\ref{fig:rhoi_gammad} that the optimization procedure described above leads to suitable coefficients ($\tilde{c}_1$, $\tilde{c}_2$, $\tilde{c}_3$) that allows $\rho^{\rm nint}_i(t)$ to follow the DDD data reasonably well.
In particular, the dislocation multiplication rates on slip systems with zero Schmid factor are well captured.
This agreement confirms the ability of the modified Kocks-Mecking relation proposed in Eqs.~(\ref{eq:KocksMeck_proposed}) and (\ref{eq:Rhoi_corr}) \nohl{to coarse-grain the dislocation multiplication data observed in DDD simulations.}
We followed this procedure to obtain $\tilde{c}$ coefficients for all of the loading orientations shown in figure~\ref{fig:stTri}a. 

\nohl{We examined the variation of the $\tilde{c}$ coefficients obtained for all of our DDD simulations in order to identify any trend in their dependence on the loading-orientation.}
Firstly, we observed a \nohl{power-law} relationship between $\tilde{c}_1$ and $\tilde{c}_2$ as follows
\begin{equation}
    \tilde{c}_2=3.3880\times 10^{4}\left(\tilde{c}_1\right)^{3/2} \,
    \label{eq:c2_c1_power_law}
\end{equation}
\nohl{can be used to estimate $\tilde{c}_2$ with $R^2=0.91$}. Subsequently, we enforced Eq.~(\ref{eq:c2_c1_power_law}) as a constraint to obtain a new set of fitting coefficients, $\{\tilde{c}_1, \tilde{c}_3\}$, for each loading orientation.
It is found that the resulting model is still in reasonable agreement with the DDD data on the predicted $\rho_i$ values.
\nohl{Considering all of the 120 simulations, applying} Eq.~(\ref{eq:c2_c1_power_law}) \nohl{results in a slight increase of the
average loss as defined by} Eq.~(\ref{eq:Loss_rhoi}), \nohl{ from $2.84\times10^{-2}\rho_0^2$ to $3.13\times10^{-2}\rho_0^2$.}
Therefore, in the following we shall adopt Eq.~(\ref{eq:c2_c1_power_law}) (for $\dot{\varepsilon} = 10^3 \, {\rm s}^{-1}$) which reduces the number of free parameters to two for dislocation multiplication in each loading-orientation.

\begin{figure}[ht]
    \begin{subfigure}[t]{0.49\textwidth} 
        \includegraphics[scale=0.135,center]{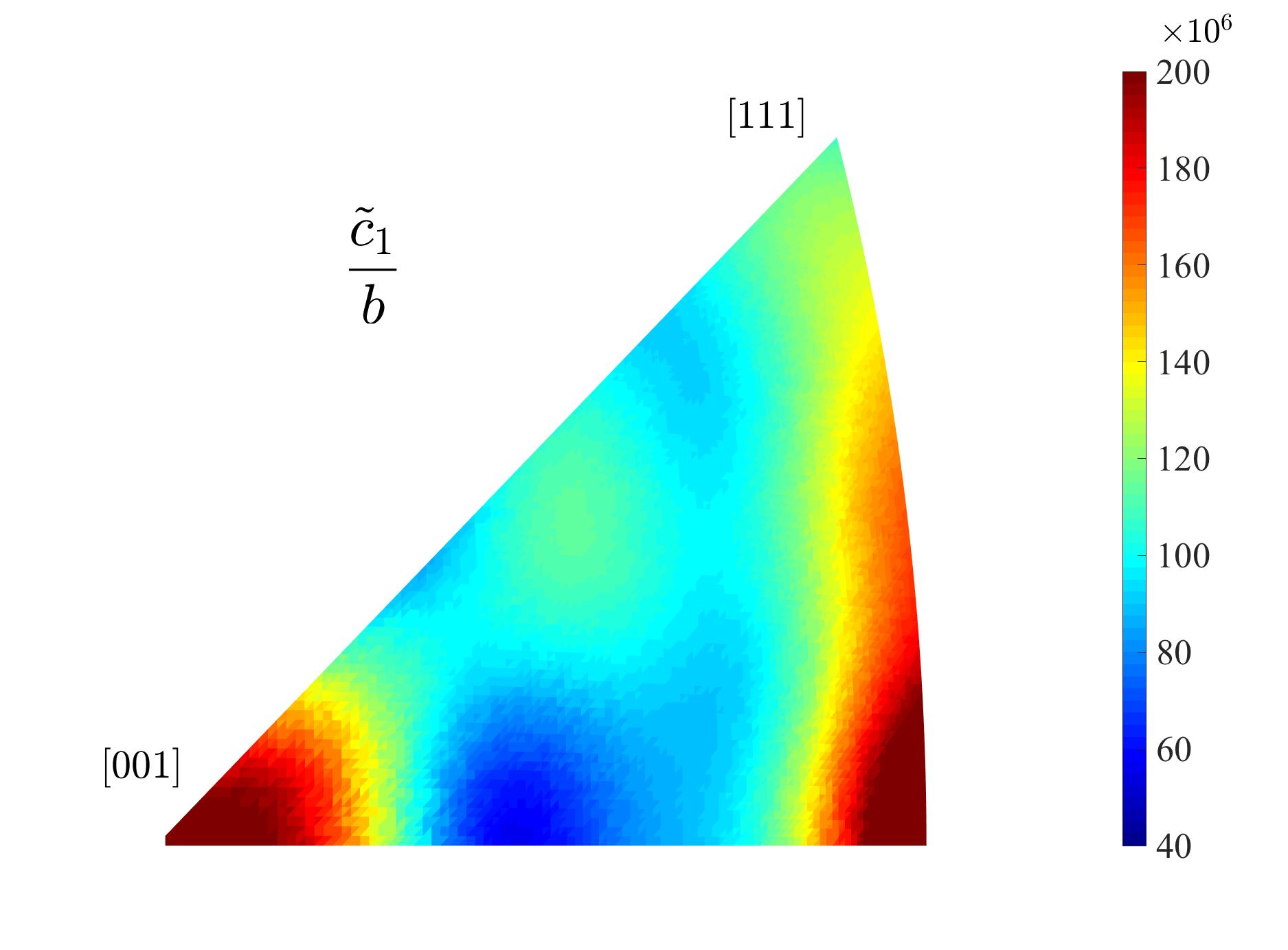}
        \caption{}
        \label{fig:C1}
    \end{subfigure}
    \begin{subfigure}[t]{0.49\textwidth} 
        \includegraphics[scale=0.135,center]{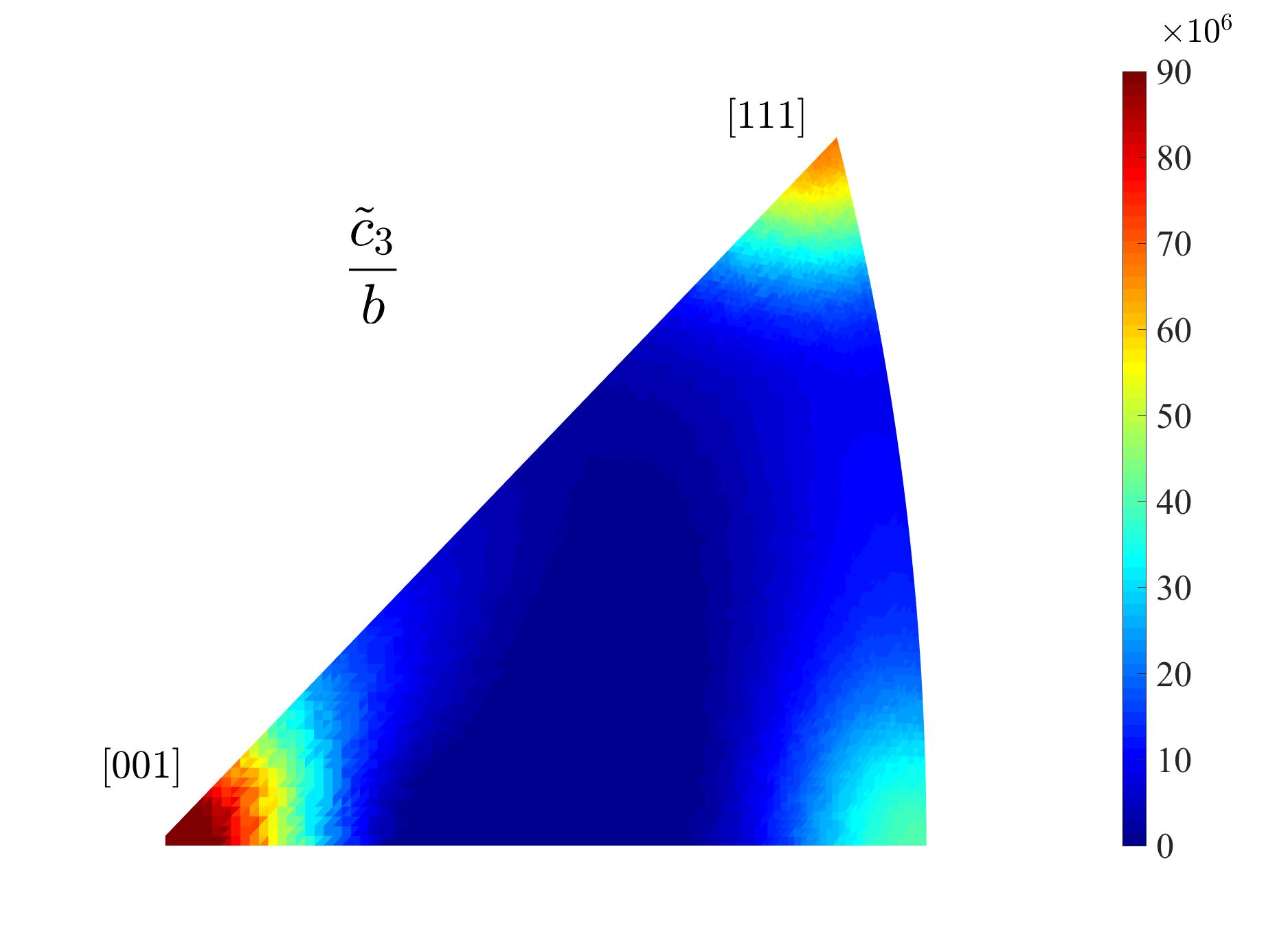}
        \caption{}
        \label{fig:C3}
    \end{subfigure}
    \caption{Smooth fields of  $\tilde{c}_1/b$ and $\tilde{c}_3/b$, obtained as {\it a posteriori} mean predicted by Gaussian Process modelling. Coefficient $\tilde{c}_2$ can be obtained from Eq.~(\ref{eq:c2_c1_power_law}).}
    \label{fig:Cfield}
\end{figure}

\nohl{Secondly, we recognized that the $\tilde{c}$ coefficients fitted to each loading orientation inevitably inherit the noise from the dislocation density data in DDD simulations, while we would expect the $\tilde{c}$ coefficients to be smooth functions of the loading orientation.}
Therefore, we smoothed the $\tilde{c}$ coefficients as functions of loading-orientation over the stereographic triangle using Gaussian Process (GP) modeling~\cite{ScikitGP, rasmussen2003gaussian}.
We observed that using the smoothed values for $\{\tilde{c}_1, \tilde{c}_3\}$ in Eqs.~(\ref{eq:KocksMeck_proposed}) and (\ref{eq:Rhoi_corr}) still leads to reasonable agreement with the DDD data on $\rho_i$ values. 
\nohl{In this case, the average of loss across all 120 simulations is $3.38\times10^{-2}\rho_0^2$.}
Figure~\ref{fig:Cfield} shows the smoothed values for $\tilde{c}_1$ and $\tilde{c}_3$, which are used in the subsequent section to construct a density-based strain-hardening model. 
\nohl{We note that $\tilde{c}_3$ is appreciable only for orientations near the three corners of the stereographic triangle, as shown in} figure~\ref{fig:Cfield}b.

\section{Strain hardening rate predicted by the constitutive model}
\label{sec:fullModel}

In Section~\ref{sec:rhoidot}, we obtained  Eqs.~(\ref{eq:KocksMeck_proposed}) and (\ref{eq:Rhoi_corr}), which can be used to predict the dislocation multiplication rate $\dot{\rho}_i$ given $\rho_i$ and the plastic shear strain rate $\dot{\gamma}_i$.
To obtain the fitting coefficients, $\tilde{c}_1$, $\tilde{c}_2$ and $\tilde{c}_3$, we made use of the plastic shear strain rate $\dot{\gamma}_i$ from the DDD data.
In Section~\ref{sec:modfTaylor}, 
we have obtained a flow rule, Eq.~(\ref{eq:vEff_expTaueff2}), which can be used to predict the plastic shear strain rate $\dot{\gamma}_i$ given the resolved shear stress $\tau_i$ and dislocation density $\rho_i$.
Therefore, we now have all the ingredients of a dislocation density-based constitutive model for strain hardening.
In this section, we use this constitutive model to predict the stress-strain curves and compare the resulting strain hardening rates from DDD simulations for uniaxial tensile deformation along different loading orientations.

For simplicity, \nohl{we adopt a visco-plastic formulation}~\cite{hutchinson1976bounds, molinari1987self, lebensohn1993self} in which we ignore the elastic strain rate (after yielding) and assume that the applied strain rate $\dot{\varepsilon}$ is entirely accommodated by the plastic shear strain rate on all the slip systems, i.e.,
\begin{equation}
    \dot{\varepsilon} = \sum_{i=1}^{12} S_i \, \dot{\gamma}_i
    \label{eq:eRate}
\end{equation}
where the Schmid factors $S_i$ are set by the loading orientation.
The resolved shear stress $\tau_i$ and plastic strain rate $\dot{\gamma}_i$ on each slip system are given by
\begin{eqnarray}
    \tau_i &=& S_i \, \sigma \label{eq:rss} \\
    \dot{\gamma}_i &=& \rho_i \, b \, v_0\exp \left[ \frac{1}{s_0}\left(\tau_i - \left(\mu b \sqrt{a'_{ij}\rho_j}-\tau_0\right) \right) \right] 
    \label{eq:gamma_i_dot_cp}
\end{eqnarray}
Combining Eqs.~\eqref{eq:eRate}, \eqref{eq:vEff_expTaueff2} and \eqref{eq:rss} yields an equation for the tensile stress $\sigma$
\begin{equation}
    \dot{\varepsilon} =  \sum_{i=1}^{12} S_i \, \rho_i \, b \, v_0\exp \left[ \frac{1}{s_0}\left(S_i\,\sigma - \left(\mu b \sqrt{a'_{ij}\rho_j}-\tau_0\right) \right) \right]
    \label{eq:implicit_sigma}
\end{equation}
where $j$ is also summed over from 1 to 12 according to Einstein's notation.
This is an implicit equation for $\sigma$, the solution of which gives the flow stress $\sigma$ as a function of the applied strain rate $\dot{\varepsilon}$ and the dislocation densities $\rho_i$, i.e., $\sigma(\dot{\varepsilon}, \{\rho_i\})$.

\nohl{We compute the quantities in our coarse-grained constitutive model as follows.}
First, we initialize the slip system densities $\rho_i$ at $t=0$ to the dislocation densities obtained from DDD in the first block $w_1$. 
Then the instantaneous flow stress $\sigma$ at time $t$ is obtained by solving Eq.~(\ref{eq:implicit_sigma}).
\nohl{In the process, the resolved shear stress and plastic shear strain rate on individual slip systems are obtained from} Eqs.~(\ref{eq:rss}) and (\ref{eq:gamma_i_dot_cp}).
Following Eqs.~(\ref{eq:KocksMeck_proposed}) and (\ref{eq:Rhoi_corr}), the dislocation multiplication rate at time $t$ can then be computed with the following relation
\begin{equation}
     \dot{\rho}_i = \dot{\gamma}_i\left(\frac{\tilde{c}_1}{b}\sqrt{a'_{ij}\rho_j}-\tilde{c}_2\,\rho_i \right) +  
                    \frac{\tilde{c}_3}{b} \left(\dot{\gamma}_{i'}\sqrt{\rho_{i''}}+ \dot{\gamma}_{i''}\sqrt{\rho_{i'}} \right)
     \label{eq:KocksMeck_corrected}
\end{equation}
Given $\dot{\rho}_i$, we can predict the dislocation density at the next time step by, e.g.,
\begin{equation}
    \rho_i(t+\Delta t) = \rho_i(t) + \dot{\rho}_i \, \Delta t \, .
\end{equation}
With the new dislocation densities, we can use Eqs.~(\ref{eq:implicit_sigma})-(\ref{eq:KocksMeck_corrected}) again to compute $\sigma$, $\tau_i$, $\dot{\gamma}_i$ and $\dot{\rho}_i$ for the next time step.  The iteration described above can be repeated to predict the stress-strain curve.

\begin{figure}[ht]
    \centering
    \begin{subfigure}[t]{0.45\textwidth} 
    \includegraphics[trim={0 0 0 13cm},clip=true,scale=0.14]{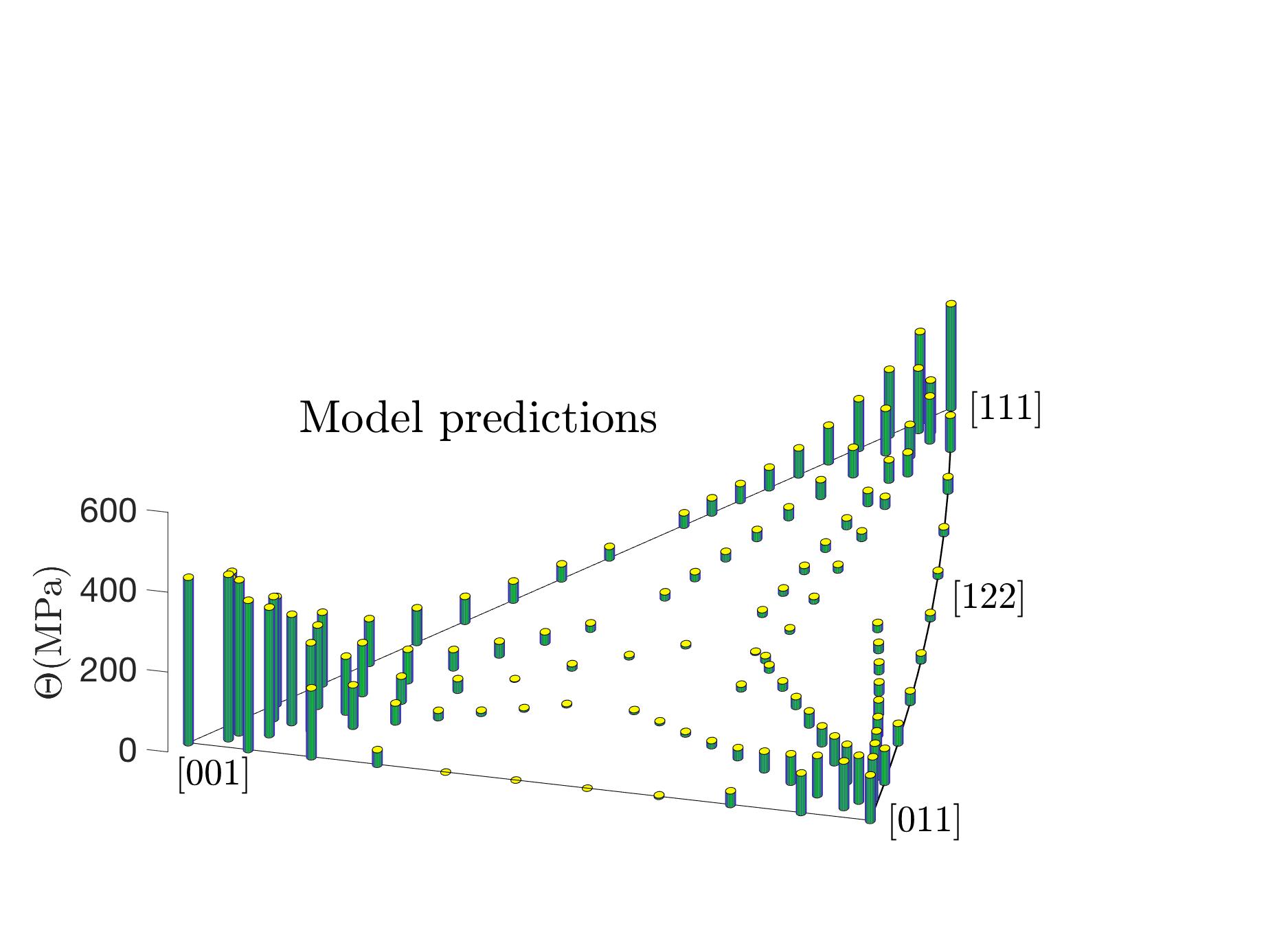}
    \caption{}
    \end{subfigure}
    \hspace{0.4in}
    \begin{subfigure}[t]{0.45\textwidth} 
     \includegraphics[trim={0 0 0 0cm},clip=true,scale=0.13,center]{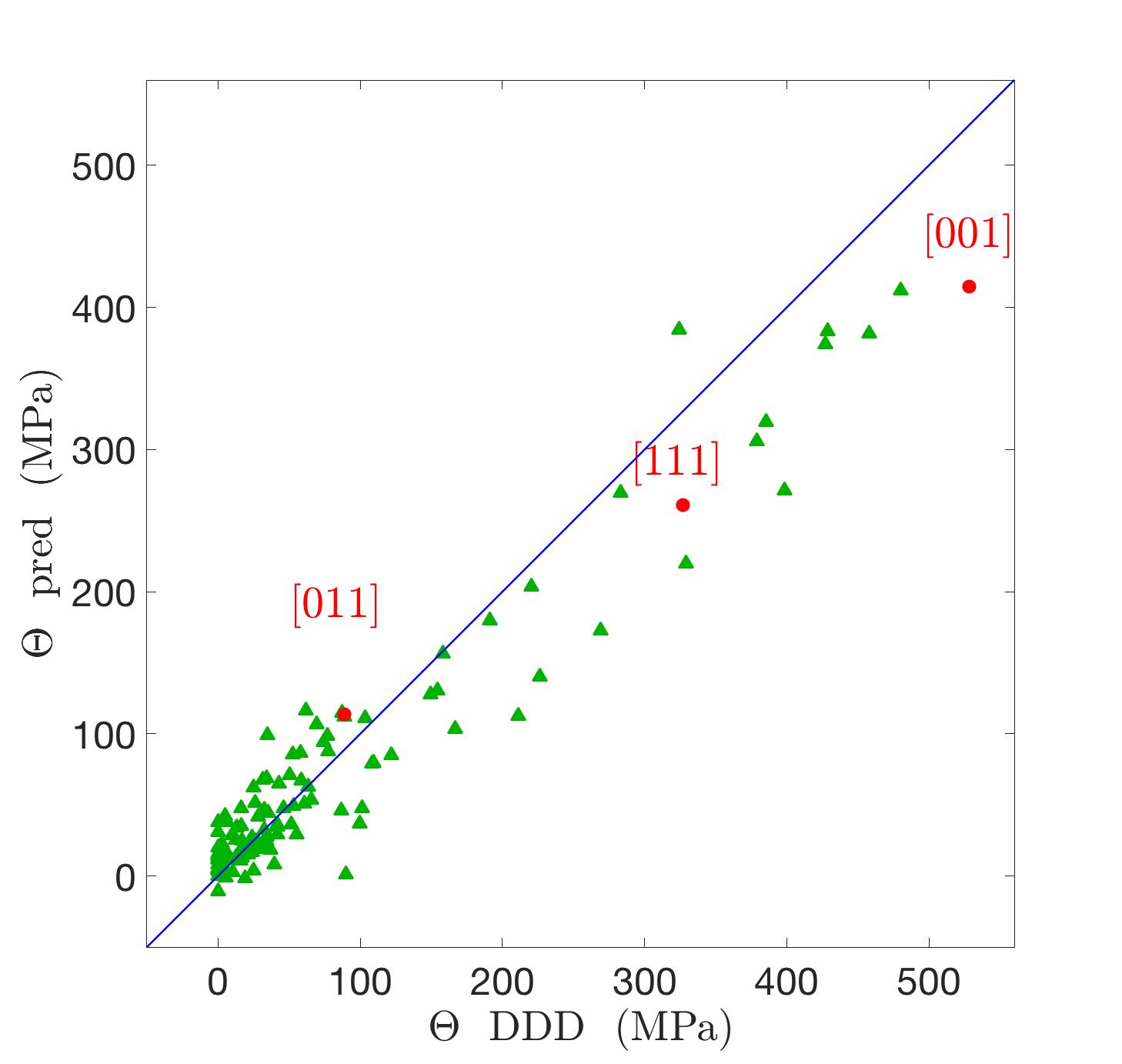}
    \caption{}
    \end{subfigure}
    \caption{Hardening rates obtained from the dislocation-based constitutive model for the same set of loading orientations in the DDD simulations. (a) Strain hardening rates predicted by the constitutive model as a function of loading orientations on the stereographic projection triangle. (b) Strain hardening rates predicted by the constitutive model compared with predictions by the DDD simulations, \nohl{with (RMSE, $R^2$) =   (36.39~MPa, 0.91)}.}
    \label{fig:stTri_fullModel}
\end{figure}\hfill

We used the constitutive model described above to predict the stress-strain curves and the strain hardening rates for all loading orientations in the DDD simulations \nohl{(at strain rate $\dot{\varepsilon} = 10^{3}\,{\rm s}^{-1}$)}.
\nohl{The flow stress predicted by the constitutive model agrees with the DDD data reasonably well, with (RMSE, $R^2$) = (0.88~MPa, 0.97) over all the blocks of all the simulations.} 
Figure~\ref{fig:stTri_fullModel}a shows the strain hardening rate predicted by the constitutive model for the same 120 orientations in the stereographic triangle as those shown in figure~\ref{fig:stTri}a.
The general dependence of strain hardening rate on loading orientation shows good agreement with the prediction from the DDD simulations shown in figure~\ref{fig:stTri}a.
The constitutive model predicts the strain hardening rate for the $[001]$, $[111]$ and $[011]$ orientation to be 464, 319 and 119~MPa, respectively, in good agreement with the DDD predictions (508, 293 and 85~MPa, respectively).
Figure~\ref{fig:stTri_fullModel}b plots the strain hardening rates predicted by the constitutive model versus those predicted by the DDD simulations.  Good agreement can be observed between the two.
We note that if the correction term to dislocation multiplication relation, Eq.~(\ref{eq:KocksMeck_proposed}) were not included, then the constitutive model would predict strain hardening rates that are significantly lower than the DDD predictions \nohl{(for orientations near the three corners of the stereographic triangle)}.
This shows that the slip-free multiplication plays an important role in the overall strain hardening behavior of the material by influencing the plastic flow and dislocation multiplication rates on active slip systems.

\section{Discussion}
\label{sec:discussion}

\subsection{Strain rate regime}

\nohl{Given that all the findings in this work are based on DDD simulations performed in the strain rates of $10^2$, $10^3$, and $10^4$\,${\rm s}^{-1}$, our conclusions are, strictly speaking, only supported by data at this high strain-rate regime.}
\nohl{However, we expect that some of the dislocation mechanisms discovered here are more general and also applicable to low strain rate loading conditions.  For example, it appears that the dependence of the strain hardening rate on the loading orientation is not very sensitive to strain rate, as shown in} Fig.~\ref{fig:stTri}.
\nohl{The separation of the flow stress expression}, Eq.~(\ref{eq:taui_logVi_aijRhoj}), \nohl{into a rate-independent and a rate-dependent term suggests that the first term is likely to remain valid under low strain-rate conditions, while the second term may be valid only in the high strain-rate conditions considered here.}
\nohl{Similarly, the exponential form of the flow rule}, Eq.~(\ref{eq:vEff_expTaueff2}), \nohl{may also be limited to the high strain-rate conditions considered here.}

\subsection{Interaction coefficients}

The modified Taylor relation as given by Eq.~(\ref{eq:taui_logVi_aijRhoj}), provides insight to understand strain hardening as a result of slip system interactions. It shows that the slip systems having higher interaction coefficients with the dominant slip system $a_{dj}$, contribute more to the hardening rate. As shown in table~\ref{table:aij}, the glissile and collinear interactions have the highest interaction coefficients.  In fact the presence of at least one such type of interaction between the dominant and one of the active slip systems was observed in all \nohl{except one} of the high hardening simulations.\footnote{\nohl{The exception is for loading along the $[3, 9, 49]$ orientation, with $\Theta = 166$ MPa.  It is a high hardening orientation but there are no glissile or collinear interactions between the dominant and other active slip systems.}}
%
%
Activation of slip systems having other types of interactions with the dominant slip system does not necessarily lead to high hardening rates, due to their lower $a_{dj}$ values.  We observed that in the low hardening simulations, there can be more than one active slip systems, but in all except one simulations, their interaction with the dominant slip systems are of the following types: coplanar, Hirth, Lomer.\footnote{\nohl{The only exception to this rule is for loading along the $[3, 3, 4]$ orientation, with $\Theta = 70$~MPa.  It is a low hardening orientation but the dominant slip system has glissile interaction with another active slip system.}} 
This evidence suggests that \nohl{glissile and collinear interactions between the dominant and other active slip systems} is the major mechanism for high strain hardening in pure, single-crystal FCC metals. Similar observations have been made in recent literature~\cite{sills2018dislocation,madec2003role}.

In the special case of $[011]$ loading orientation which is categorized as low-hardening, 4 active slip systems have identical Schmid factors.  Yet the interactions between the dominant system  and the other three active systems are of the types: coplanar, Hirth and Lomer, none of which has high enough $a_{dj}$ value to cause high hardening rate. This is true for both sets of the interaction coefficients presented in table~\ref{table:aij}. 
Note that in the present work we adopted the new set of interaction coefficients $a'_{ij}$, such that consistent values for $\tau_0$ and $v_0$ can be used in both Eq.~(\ref{eq:vEff_expTaueff}) and Eq.~(\ref{eq:vEff_expTaueff2}). This is not possible if the coefficients $a_{ij}$ calculated from~\cite{kubin2008modeling} were used.

Dislocations segments that do not belong to one of the $\frac{1}{2}\langle110\rangle\{111\}$ slip systems are immobile segments (e.g., Hirth or Lomer junctions), and they constitute about 25\% of the total dislocation density. 
Despite their considerable fraction, the modified Taylor relation as given by Eq.~(\ref{eq:taui_logVi_aijRhoj}) only includes $\frac{1}{2}\langle110\rangle\{111\}$ dislocations; yet good agreement with the DDD data was still obtained. 
One possible explanation is that within the strain range of our DDD simulations, the density ratio between junction dislocations and mobile dislocations stay more or less constant, so that the effect of the junction dislocation density may be absorbed in the interaction coefficients $a'_{ij}$. 
It is possible for the fraction of dislocation junction density to further increase at higher strains, in which case the effect of junction density may need to be accounted for explicitly.

\subsection{Glissile junction mechanisms for multiplication}
\label{sec:glissilejunction}
A key component of our proposed continuum model is a modified Kocks-Mecking expression to describe the evolution of the $\rho_i$. In order to account for the dislocation density increase on those slip systems with negligible plastic activity \nohl{(slip-free multiplication)}, we proposed to add a correction term as given by Eq.~(\ref{eq:Rhoi_corr}).  
The dislocation density increase on zero-Schmidt-factor slip systems has been observed experimentally~\cite{higashida1986formation}.
This type of dislocation multiplication was later reported in the DDD simulations~\cite{weygand2014mechanics, stricker2015dislocation}, where it was attributed to the glissile junction formation, \nohl{and more recently in ultra-scale Molecular Dynamics simulations of crystal plasticity}~\cite{zepedaruiz2019metal}.
A multiplication rate formulation based on the glissile junction mechanism was proposed in~\cite{stricker2015dislocation}, which in our notation \nohl{could} be expressed as follows 
\begin{equation}
    \dot{\rho}_{i}^{\rm corr}(\{\dot{\gamma}_j\}, \{\rho_j\}) 
    = \sum_{(j,k)}\frac{\tilde{c}'_3}{b} \left(\dot{\gamma}_{j}\sqrt{\rho_{k}}+ \dot{\gamma}_{k}\sqrt{\rho_{j}} \right)
    \label{eq:Rhoi_corr_glissile}
\end{equation}
where a pair of slip systems, $j$ and $k$, can react and form glissile junctions, which is a mobile dislocation on slip system $i$. 
In this work, we find that the correction term, Eq.~({\ref{eq:Rhoi_corr}}), based on the coplanar interaction, leads to predictions of $\rho_i$ time histories that are in better agreement with our DDD simulation data.
\nohl{In order to quantify the error, loss function as shown by} Eq.~(\ref{eq:Loss_rhoi}) can be used. Average loss value across all 120 simulations that have been used for fitting $\rho_i$ using Eq.~{\eqref{eq:Rhoi_corr}} is $2.84\times 10^{-2}\,\rho_0^2$,
\nohl{where $\rho_0$ is the initial total dislocation density. The corresponding loss value increases by more than $50\%$ to $5.27 \times 10^{-2} \rho_0^2$ when Eq.~{\eqref{eq:Rhoi_corr_glissile}} is used. Moreover, excluding the correction term in Eq.~{\eqref{eq:KocksMeck_proposed}}, i.e., $\dot{\rho}_i^{\rm corr}=0$, leads to a high average loss of $7.53 \times 10^{-2} \rho_0^2$. Note that mentioned loss values are associated with the result of fitting Eq.~{\eqref{eq:KocksMeck_proposed}} to the DDD data, prior to applying the Eq.~{\eqref{eq:c2_c1_power_law}} on $\tilde{c}_1$ and $\tilde{c}_2$, or Gaussian process modeling.}

\nohl{In this work, by choosing the correction term,} Eq.~({\ref{eq:Rhoi_corr}}), \nohl{based on the coplanar interaction, we do not imply that the glissile junction is not important for dislocation multiplication.  In fact, glissile junctions form abundantly in our DDD simulations.  However, it is possible that their contribution to dislocation multiplication has been effectively accounted for in the first term of} Eq.~(\ref{eq:KocksMeck_proposed}), \nohl{so that adding the correction term}, Eq.~(\ref{eq:Rhoi_corr_glissile}), \nohl{does not lead to a significant reduction of the fitting error.}

We point out that an ideal evolution model for $\dot{\rho}_i$ should not \nohl{have an explicit dependence} on the loading orientation. This is not the case for Eqs.~(\ref{eq:KocksMeck_proposed}) and (\ref{eq:Rhoi_corr}) where the coefficients $\tilde{c}$ are loading-orientation dependent. 
This is also the case for most of the equations proposed in literature, e.g.~\cite{devincre2008dislocation, stricker2015dislocation}. 
This means Eqs.~(\ref{eq:KocksMeck_proposed}) and (\ref{eq:Rhoi_corr}), even though more consistent with the DDD data than their existing counterparts, \nohl{are still not in the most desirable form for describing dislocation multiplication in a CP model.}
More investigation is needed to extract a dislocation multiplication model that is more physics-based and generalizable.

\subsection{Cross slip}
\label{sec:crossslip}

Cross slip has been suppressed in all of our DDD simulations in this work.  We believe this is a reasonable approximation for the present purpose for the following reasons.
In this work we focus on the initial strain hardening rate in the regime of less than $5\%$ plastic shear strain.  On the other hand, cross-slip is expected to play a significant role in the so-called dynamic recovery regime, which becomes prominent at much higher strains marking the transition of strain hardening from Stage II to Stage III.
We have also observed that the predictions of the initial strain hardening rate by the DDD simulations without accounting for cross-slip are in good qualitative agreement with existing experiments, especially in the orientation dependence of strain hardening rate.
Therefore, mechanisms other than cross slip must be playing a dominant role, and cross slip can be considered as a perturbation, in the orientation dependence of the strain hardening rate, which is the focus of this work.
We expect that in the simulations where cross-slip is activated, Eqs.~(\ref{eq:taui_logVi_aijRhoj}) and (\ref{eq:taud_mu_b_over_ld_logVd}) would still be valid, but with possibly different fitting coefficients, while Eq.~(\ref{eq:Rhoi_corr}) \nohl{may need} an additional correction term to account for multiplication due to the cross-slip mechanism. Example of such correction term is given in~\cite{sudmanns2019dislocation}.

There are still unresolved controversies on cross slip of screw dislocations in FCC metals and how it should be modeled in DDD simulations.
The cross slip rate is known to be influenced by many stress components, such as the Schmid stress on the cross slip plane, Escaig stress on the cross slip plane, and Escaig stress on the original slip plane~\cite{kang2014stress}.
There are some controversies on whether or not the Schmid stress on the original slip plane should be accounted for in calculating the cross slip rate.
Recently, it has been shown that the cross slip rate can be greatly enhanced at dislocation intersections~\cite{hussein2015microstructurally}.  Because describing the rate of intersection cross-slip involves many parameters, it appears that more research is needed before a consensus emerges on the best way to account for intersection cross-slip in DDD simulations.
In this context, our work provides an example of a systematic data-driven approach for constructing crystal plasticity models by coarse-graining the DDD model.
As a first demonstration, the DDD model is chosen to be simple and completely specified by a few parameters (see Section~\ref{sec:DDDsetup}).
In the future, the same coarse-graining procedure can be applied to coarse-grain more complex DDD models, e.g., with different kinds of cross slip models enabled.  
Comparing the resulting crystal plasticity models with the one obtained in this work would reveal the effect of cross slip at the macroscopic scale.

\subsection{Future developments of density-based models}

For simplicity, the constitutive model considered here only contains one microstructural state variable per slip system, the dislocation density $\rho_i$.
In the future, more state variables may be considered to provide a more refined description of the dislocation microstructure.
For example, motivated by~\cite{sills2018dislocation}, one may consider two microstructural state variables per slip system, $\rho_i$ and $N_i$, where $N_i$ is the number of dislocation links on slip system $i$.
We note that the average length of dislocation links can be expressed as $\bar{l}_i = \rho_i / N_i$, and that a non-dimensional parameter, $\phi_i = N_i^2/\rho_i^3$ may be defined to characterize the structure of dislocations on slip system $i$.
Treating $\rho_i$ and $N_i$ as independent state variables implies that the non-dimensional parameter $\phi_i$ is not the same across all slip systems.  This has been confirmed by our DDD data and will be further discussed in a subsequent publication.
If one were to construct such a (more complex) constitutive model based on $\rho_i$ and $N_i$, provided that sufficient DDD data is available, then in addition to the flow rule $\dot{\gamma}_i$ and the multiplication rate $\dot{\rho}_i$, a relation must be found for describing the rate of link number increase $\dot{N}_i$ based on DDD data.  These possibilities will be explored in future studies.

We note that in the analyses above, relatively simple ``data-mining'' techniques were employed\nohl{. Though more advanced tools were used to obtain the fitting coefficients once a functional form is chosen, the main results}~(e.g. Eq.~(\ref{eq:vEff_expTaueff}) and \nohl{ slip-free multiplication) were obtained} mostly via manual manipulation and plotting of the data.
Even with such simple techniques, a great deal of insight could be gained. In future work, we hope to leverage more advanced data science and machine learning techniques to learn even more from our ever expanding DDD database.

\section{Conclusion}

We presented a continuum model of strain hardening by systematically coarse-graining a large set of DDD simulation data.
More than 200 total DDD simulations was performed using two different initial configurations, along 120 loading orientations, subjected to three strain rates of $10^2~\rm s^{-1},~10^3~\rm s^{-1}$ and $10^4~\rm s^{-1}$. 
The resulting crystal plasticity model uses the dislocation densities $\rho_i$ on 12 slip systems as the microstructural state variables.
The constitutive relations consist of a plastic flow rule that combines the generalized Taylor relation and the Orowan relation, and a modified Kocks-Mecking relation for dislocation multiplication.

In the process of developing the continuum model, several important aspects of the physics of dislocation interactions were revealed. 
First, we found a simple linear relationship between the resolved shear stress, average link length, and logarithm of the plastic shear strain rate on the dominant slip system across all loading orientations.
This relation can be rewritten as a flow rule to predict the plastic shear strain rate as an exponential function of the resolved shear stress, and is found to be well supported by DDD data on all slip systems.
Second, we found that the dislocation multiplication rate of a slip system depends, not only on the plastic strain rate of its own, but also on the other two slip systems sharing the same slip plane. 
We found that a correction term, when added to the generalized Kocks-Mecking model, can capture the dislocation multiplication rates on inactive slip systems that are consistent with the DDD data.
The constructed crystal plasticity model successfully predicts strain hardening rates that are consistent with the DDD data, which are also in qualitative agreement with existing experiments.

\vspace{25px}

\noindent {\bf Acknowledgement}
This work was supported by the U.S. Department of Energy, Office of Basic Energy Sciences, Division of Materials Sciences and Engineering under Award No. DE-SC0010412 (Sh. A. and W. C.).
Part of the DDD simulations were performed using allocation MSS190011 on the SDSC Comet-GPU and PSC Bridges-GPU clusters of the Extreme Science and Engineering Discovery Environment (XSEDE), which was supported by National Science Foundation grant number ACI-1548562.
Part of NB's work was performed under the auspices of the U.S. Department of Energy by Lawrence Livermore National Laboratory under contract DE-AC52-07NA27344.

\begin{appendices}
\section{Analysis of power-law flow rule}
\label{appx:powerLaw}
\counterwithin{figure}{section}
\setcounter{figure}{0}  
\counterwithin{equation}{section}
\setcounter{equation}{0}  
\counterwithin{table}{section}
\setcounter{table}{0}

\begin{figure}[ht]
    \centering
    \begin{subfigure}[t]{0.45\textwidth} 
    \includegraphics[trim={1cm 0 0 0cm},clip=true,scale=0.18]{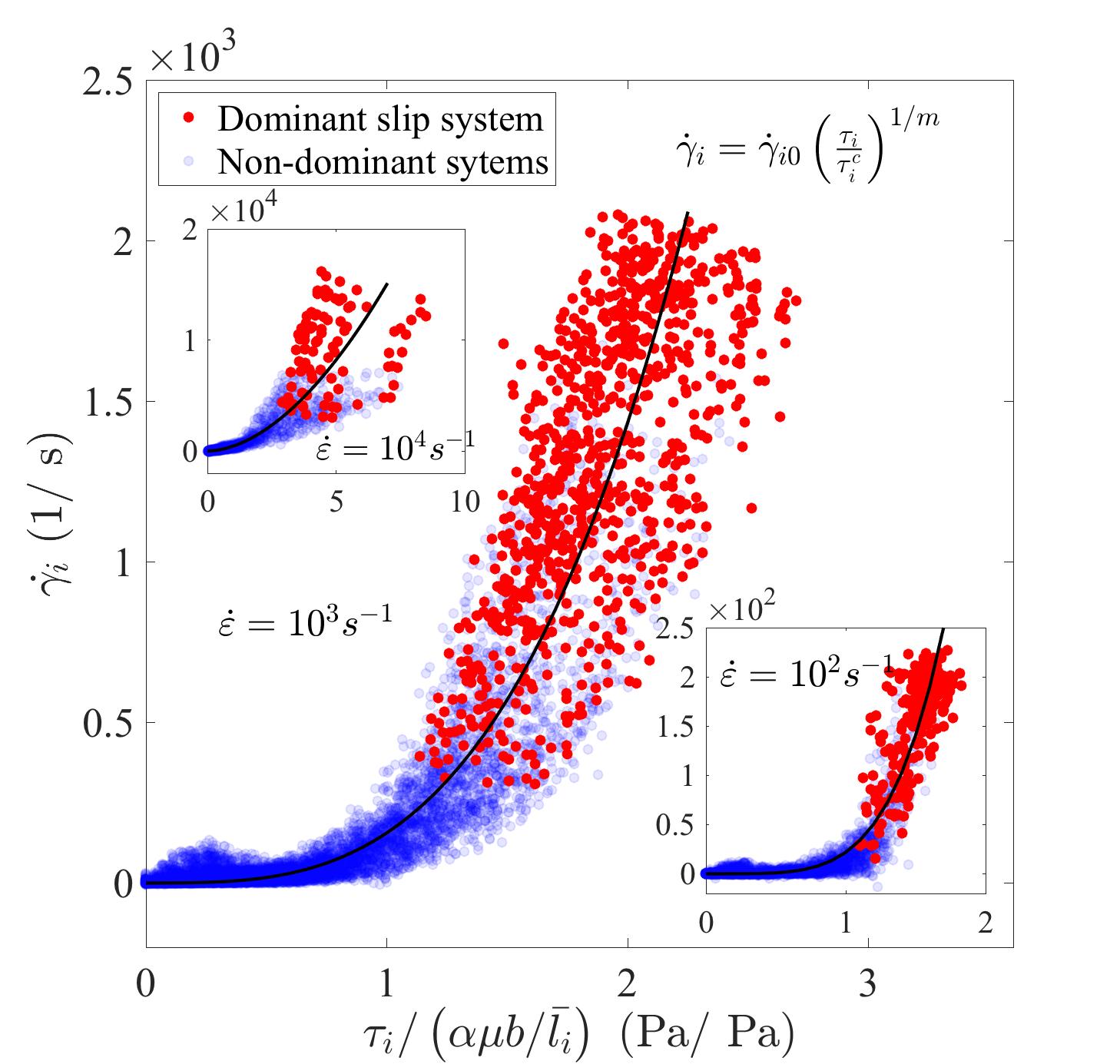}
    \caption{}
    \end{subfigure} \hfill
    \hspace{0.4in}
    \begin{subfigure}[t]{0.45\textwidth} 
     \includegraphics[trim={0 0 0 0cm},clip=true,scale=0.18,center]{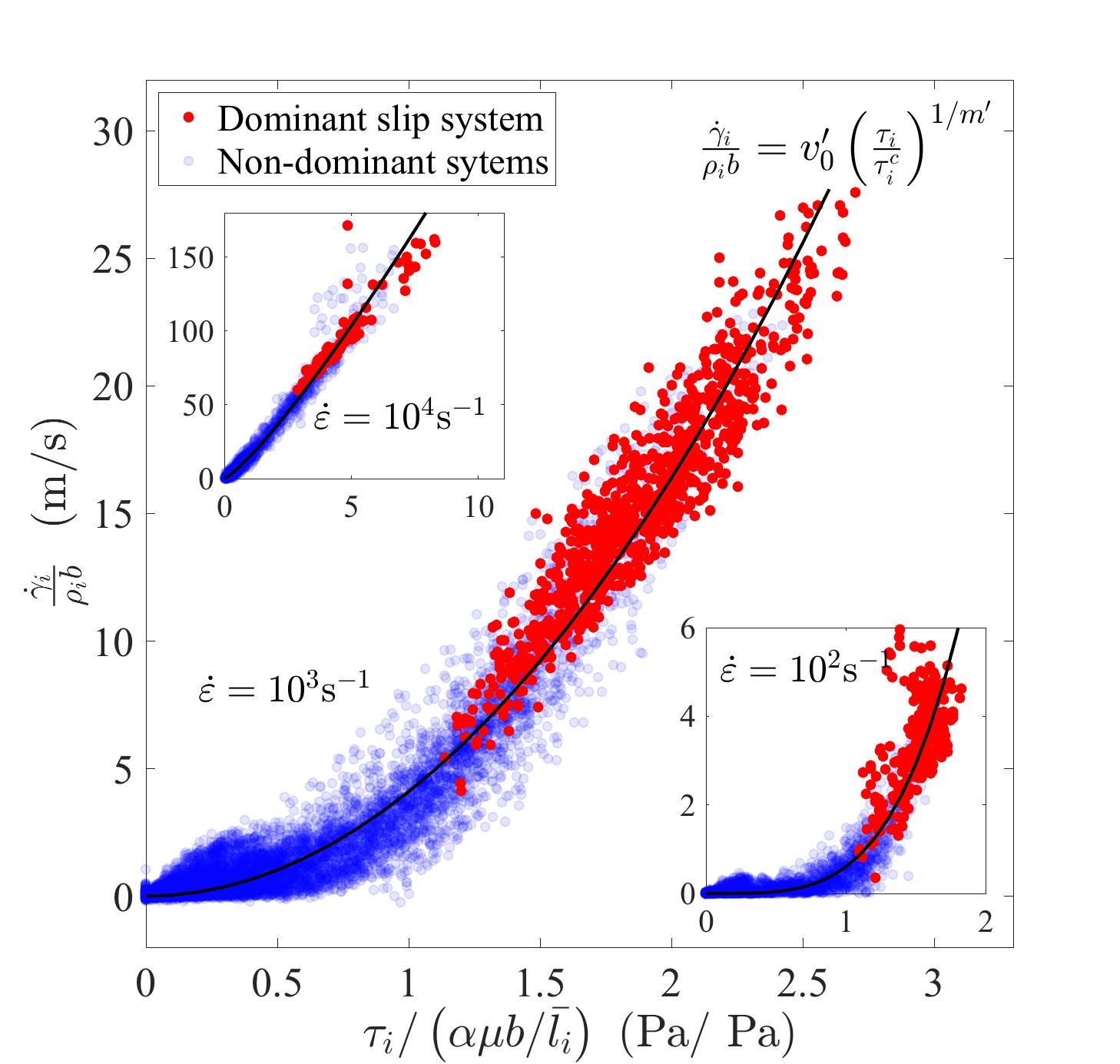}
    \caption{}
    \end{subfigure}
    \caption{ Power-law relationship between (a) ${\dot{\gamma}_i}$ and $\tau_i / \tau_i^c$  as stated in Eq.~(\ref{eq:powerlaw}) (solid line), and (b) ${\dot{\gamma}_i}/({\rho_i b})$ and $\tau_i / \tau_i^c$ as stated in Eq.~(\ref{eq:powerLaw_vel}), with $\tau_i^c=\alpha\mu b/\bar{l}_i$. \nohl{The error between the fitted curve and DDD data is (a) (RMSE, $R^2$) =  (1.7$\times10^{-2}~\rm{1/s}$, 0.87),  (1.89$\times10^{2}~\rm{1/s}$, 0.79), and  (2.1$\times10^{3}~\rm{1/s}$, 0.47) and (b) (RMSE, $R^2$) = (0.36~$\rm{m/s}$, 0.88),  (1.04~$\rm{m/s}$, 0.96), and  (7.65~$\rm{m/s}$, 0.96) for $\dot{\varepsilon} = 10^2\,\rm{s}^{-1}, 10^3\,\rm{s}^{-1}$ and $10^4\,\rm{s}^{-1}$, respectively.} Dots are data points correspond to the slip systems of total of $\approx$ 170 DDD simulations, from two different initial configurations, under strain rate $\dot{\varepsilon}=10^3\,{\rm s^{-1}}$. The insets contain data points from 54 and 18 DDD simulations under strain rate $\dot{\varepsilon}=10^2\,{\rm s^{-1}}$ and $10^4\,{\rm s^{-1}}$, respectively. 
    }
    \label{fig:PowerLaw}
\end{figure}

Power-law flow rule as given in Eq.~(\ref{eq:powerlaw}) is one of the most commonly used flow rules in the literature \cite{roters2010overview}. In this work, we proposed an exponential flow rule which better represents the DDD data. In this appendix we provide evidence of why the exponential flow rule is more consistent with our DDD data than the power-law.  Eq.~(\ref{eq:powerlaw}) is reproduced below:

\begin{equation*}
    \dot{\gamma}_i = \dot{\gamma}_0 \left( \frac{\tau_i}{\tau_i^{\rm c}} \right)^{1/m}
\end{equation*}
where $\dot{\gamma}_{0}$ and $m$ are material specific coefficients which, in the quasi static loading condition, are typically in ranges of $10^{-16}-10^{-3}\,\rm{s^{-1}}$, and $0.005-0.01$, respectively \cite{groh2009multiscale,shanthraj2011dislocation, demir2016physically}. 
The critical resolved shear stress, $\tau_i^c$, which characterizes the onset of dislocation glide on the slip system $i$, is commonly prescribed phenomenologically, or estimated with Taylor-like relations in dislocation-density based models, e.g., using (the first term on the right hand side of) Eq.~(\ref{eq:extended-Taylor}) \cite{shanthraj2011dislocation,fivel1998identification,devincre2015physically}.
Here to compare with the exponential flow rule, Eq.~(\ref{eq:vEff_expTaueff}), the critical resolved shear stress in the power-law flow rule is set to $\tau_i^c=\alpha \mu b / \bar{l}_i$.
The quality of the fit of Eq.~(\ref{eq:powerlaw}) to the DDD data is poor, as shown in Fig.~\ref{fig:PowerLaw}(a).  A wide scatter of the data points is observed especially for higher strain rates of $10^3\,{\rm s}^{-1}$ and $10^4\,{\rm s}^{-1}$.

\begin{table}[ht] 
\centering
\small
\caption{Fitting coefficients in the proposed flow rule shown in Eqs.~(\ref{eq:powerlaw}) and (\ref{eq:powerLaw_vel}).
} \label{table:powerLaw}
\begin{tabular}{ P{3cm}  P{1.2cm}  P{1.2cm} P{1.5cm} P{1.5cm}  P{1.2cm} }
\toprule
  & $\dot{\gamma}_{0}$ & $m$ & $v'_0\,\rm(m\cdot s^{-1})$ & $m'$  \\
\toprule
$\dot{\varepsilon}=10^2$\,s$^{-1}$  & 21.8 & 0.22 & 0.6 & 0.26 \\
\midrule
$\dot{\varepsilon}=10^3$\,s$^{-1}$  & 156.0 & 0.31 & 4.1 & 0.50  \\
\midrule
$\dot{\varepsilon}=10^4$\,s$^{-1}$ & 438.0 & 0.55 & 15.0 & 0.83   \\
\bottomrule
\end{tabular}
\end{table}

We notice that by using the Orowan relation, $\dot{\gamma}_i = \rho_i b \bar{v}_i$, and expressing the average velocity $\bar{v}_i$ as a power-law leads to a much better agreement with the DDD data.
This leads to the following modified power-law flow rule
\begin{equation}
    \frac{\dot{\gamma}_i}{\rho_i b} = v'_0\left(  \frac{\tau_i}{\tau_i^c} \right)^{1/m'}.
    \label{eq:powerLaw_vel}
\end{equation}
Fig.~\ref{fig:PowerLaw}(b) shows the level of agreement of Eq.~(\ref{eq:powerLaw_vel}) with the DDD data, which is comparable with the exponential equation shown in Fig.~\ref{fig:exponential_flow_rule}(b). 
The average error shown in Fig.~\ref{fig:PowerLaw}(b) is RMSE = $1.05 \,{\rm m/s}$ for the strain rate of $\dot{\varepsilon}=10^3\,\rm{s^{-1}}$, which is even slightly less than the corresponding error in Fig.~\ref{fig:exponential_flow_rule}(a).

\hfill

\begin{figure}[ht]
    \centering
    \includegraphics[trim={0 0 0 0cm},clip=true,scale=0.19,center,valign=t]{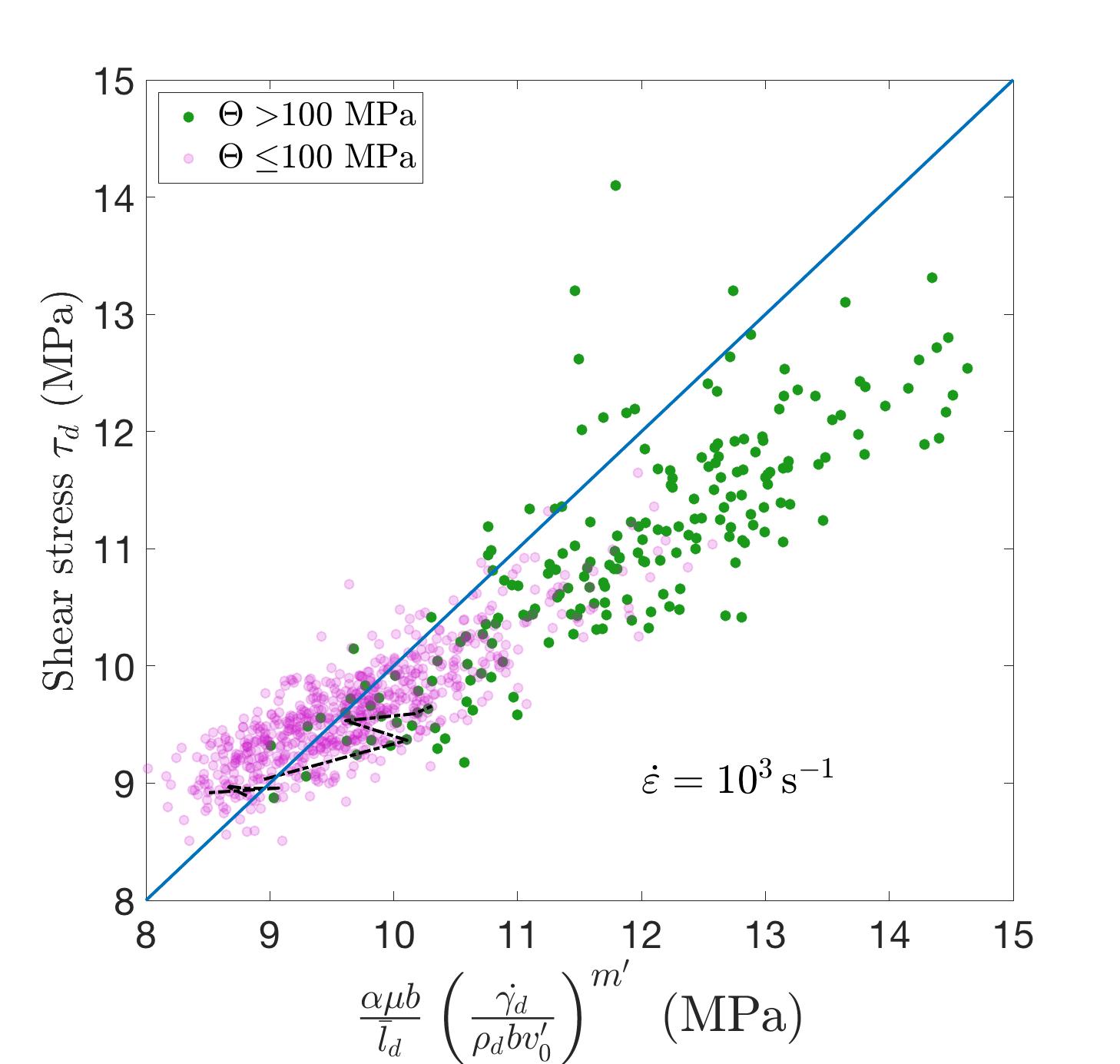}
    \caption{
    $\tau_d$ from the DDD simulations compared with the prediction from the inverse of power-law flow rule as stated in Eq.~(\ref{eq:invPowerLaw_vel}), with 
    \nohl{(RMSE, $R^2$) = ($0.71~\rm{MPa}$, 0.36).}
    }
    \label{fig:invPowerLaw_Vel}
\end{figure}

Thus it appears that both Eq.~(\ref{eq:powerLaw_vel}) and Eq.~(\ref{eq:vEff_expTaueff}) are in reasonable agreement with the DDD data.
However, the difference between the two models becomes apparent when the flow rule is rewritten as an expression for the flow stress given the shear strain rate.
For the exponential flow rule, Eq.~(\ref{eq:vEff_expTaueff}), the corresponding flow stress expression is Eq.~(\ref{eq:taud_mu_b_over_ld_logVd}), which is in very good agreement with the DDD data (RMSE = 0.34~MPa), as shown in Fig.~\ref{fig:taud_one_over_ld_B}. 
For the modified power-law flow rule, the corresponding flow stress expression is as follows
\begin{equation}
    \tau_d = \left( \frac{\alpha\mu b}{\bar{l}_d} \right) \cdot \left(\frac{\dot{\gamma}_d}{\rho_d b v'_0} \right)^{m'}.
    \label{eq:invPowerLaw_vel}
\end{equation}
Comparison of Eq.~(\ref{eq:invPowerLaw_vel}) with the DDD data is presented in the Fig.~\ref{fig:invPowerLaw_Vel}(a). 
The agreement between Eq.~(\ref{eq:invPowerLaw_vel}) and the DDD data is poor, with a high error of RMSE = 0.71~MPa.
It can be seen that, especially for the low-hardening cases, the multiplicative factor, $\left(\frac{\dot{\gamma}_d}{\rho_d b v_0'}\right)^{m'}$, effectively modifies the Taylor coefficient $\alpha$.  
This modification lowers the slope of the Taylor plot (i.e., $\tau_d$ vs $\alpha \mu b/\bar{l}_d$) and is inconsistent with the DDD data.
The DDD data strongly suggest that the strain rate correction to the Taylor relation should be additive, leaving the Taylor factor $\alpha$ unchanged, as in Eq.~(\ref{eq:taud_alpha_mu_b_over_ld}).

If the power-law flow rule Eq.~(\ref{eq:powerLaw_vel}) is used in the constitutive model described in section~(\ref{sec:fullModel}), in place of the exponential flow rule, we observe that the constitutive model systematically predicts a higher flow stress than DDD.  The average error in the flow stress over all the blocks of all the simulations is \nohl{(RMSE, $R^2$) = (2.16~MPa, 0.81)},
i.e., \nohl{RMSE is} more than a factor of 2 higher than for the exponential flow rule (RMSE = 0.88~MPa). 
Given the results presented above, we conclude that the exponential flow rule is more consistent with our DDD data than the (modified) power-law flow rule.

\end{appendices}


\medskip

\bibliographystyle{ieeetr}
\bibliography{refs}

\newcommand{\noop}[1]{}
\begin{thebibliography}{10}

\bibitem{bulatov2006computer}
V.~Bulatov and W.~Cai, {\em Computer simulations of dislocations}, vol.~3.
\newblock Oxford University Press on Demand, 2006.

\bibitem{arsenlis2007enabling}
A.~Arsenlis, W.~Cai, M.~Tang, M.~Rhee, T.~Oppelstrup, G.~Hommes, T.~G. Pierce,
  and V.~V. Bulatov, ``Enabling strain hardening simulations with dislocation
  dynamics,'' {\em Modelling and Simulation in Materials Science and
  Engineering}, vol.~15, no.~6, p.~553, 2007.

\bibitem{rao2008athermal}
S.~I. Rao, D.~Dimiduk, T.~A. Parthasarathy, M.~Uchic, M.~Tang, and C.~Woodward,
  ``Athermal mechanisms of size-dependent crystal flow gleaned from
  three-dimensional discrete dislocation simulations,'' {\em Acta Materialia},
  vol.~56, no.~13, pp.~3245--3259, 2008.

\bibitem{ARMRnbertin}
N.~Bertin, R.~B. Sills, and W.~Cai, ``Frontiers in the simulation of
  dislocations,'' {\em Annual Review of Materials Research}, \noop{3001}in
  press, 2020.

\bibitem{roters2010overview}
F.~Roters, P.~Eisenlohr, L.~Hantcherli, D.~D. Tjahjanto, T.~R. Bieler, and
  D.~Raabe, ``Overview of constitutive laws, kinematics, homogenization and
  multiscale methods in crystal plasticity finite-element modeling: Theory,
  experiments, applications,'' {\em Acta Materialia}, vol.~58, no.~4,
  pp.~1152--1211, 2010.

\bibitem{hutchinson1976bounds}
J.~W. Hutchinson, ``Bounds and self-consistent estimates for creep of
  polycrystalline materials,'' {\em Proceedings of the Royal Society of London.
  A. Mathematical and Physical Sciences}, vol.~348, no.~1652, pp.~101--127,
  1976.

\bibitem{peirce1983material}
D.~Peirce, R.~J. Asaro, and A.~Needleman, ``Material rate dependence and
  localized deformation in crystalline solids,'' {\em Acta metallurgica},
  vol.~31, no.~12, pp.~1951--1976, 1983.

\bibitem{franccois2012mechanical}
D.~Fran{\c{c}}ois, A.~Pineau, and A.~Zaoui, {\em Mechanical Behaviour of
  Materials: Volume 1: Micro-and Macroscopic Constitutive Behaviour}, vol.~180.
\newblock Springer Science \& Business Media, 2012.

\bibitem{kocks1976laws}
U.~Kocks, ``Laws for work-hardening and low-temperature creep,'' 1976.

\bibitem{becker1991analysis}
R.~Becker, ``Analysis of texture evolution in channel die compression—i.
  effects of grain interaction,'' {\em Acta metallurgica et materialia},
  vol.~39, no.~6, pp.~1211--1230, 1991.

\bibitem{johnson1983constitutive}
G.~R. Johnson and W.~H. Cook, ``A constitutive model and data for metals
  subjected to large strains, high strain rates and high temperatures,'' in
  {\em Proceedings of the 7th International Symposium on Ballistics}, vol.~21,
  pp.~541--547, The Netherlands, 1983.

\bibitem{roters2019damask}
F.~Roters, M.~Diehl, P.~Shanthraj, P.~Eisenlohr, C.~Reuber, S.~L. Wong,
  T.~Maiti, A.~Ebrahimi, T.~Hochrainer, H.-O. Fabritius, {\em et~al.},
  ``Damask--the d{\"u}sseldorf advanced material simulation kit for modeling
  multi-physics crystal plasticity, thermal, and damage phenomena from the
  single crystal up to the component scale,'' {\em Computational Materials
  Science}, vol.~158, pp.~420--478, 2019.

\bibitem{MSMSE2020}
E.~V. der Giessen, P.~Schultz, N.~Bertin, V.~Bulatov, W.~Cai, G.~Csanyi,
  S.~Foiles, M.~Geers, C.~Gonzale, M.~Huetter, W.~Kim, D.~Kochmann, J.~LLorca,
  A.~Mattsson, J.~Rottler, A.~Shluger, R.~Sills, I.~Steinbach, A.~Strachan, and
  E.~Tadmor, ``Roadmap on multiscale materials modeling,'' {\em Modelling and
  Simulation in Materials Science and Engineering}, 2020.

\bibitem{ma2004constitutive}
A.~Ma and F.~Roters, ``A constitutive model for fcc single crystals based on
  dislocation densities and its application to uniaxial compression of
  aluminium single crystals,'' {\em Acta materialia}, vol.~52, no.~12,
  pp.~3603--3612, 2004.

\bibitem{groh2009multiscale}
S.~Groh, E.~Marin, M.~Horstemeyer, and H.~M. Zbib, ``Multiscale modeling of the
  plasticity in an aluminum single crystal,'' {\em International Journal of
  Plasticity}, vol.~25, no.~8, pp.~1456--1473, 2009.

\bibitem{chandra2015multiscale}
S.~Chandra, M.~Samal, V.~Chavan, and R.~Patel, ``Multiscale modeling of
  plasticity in a copper single crystal deformed at high strain rates,'' {\em
  Plasticity and Mechanics of Defects}, vol.~1, no.~1, 2015.

\bibitem{pinna2015assessment}
C.~Pinna, Y.~Lan, M.~Kiu, P.~Efthymiadis, M.~Lopez-Pedrosa, and D.~Farrugia,
  ``Assessment of crystal plasticity finite element simulations of the hot
  deformation of metals from local strain and orientation measurements,'' {\em
  International Journal of Plasticity}, vol.~73, pp.~24--38, 2015.

\bibitem{mello2016effect}
A.~W. Mello, A.~Nicolas, R.~A. Lebensohn, and M.~D. Sangid, ``Effect of
  microstructure on strain localization in a 7050 aluminum alloy: comparison of
  experiments and modeling for various textures,'' {\em Materials Science and
  Engineering: A}, vol.~661, pp.~187--197, 2016.

\bibitem{sills2016advanced}
R.~B. Sills, A.~Aghaei, and W.~Cai, ``Advanced time integration algorithms for
  dislocation dynamics simulations of work hardening,'' {\em Modelling and
  Simulation in Materials Science and Engineering}, vol.~24, no.~4, p.~045019,
  2016.

\bibitem{bertin2019gpu}
N.~Bertin, S.~Aubry, A.~Arsenlis, and W.~Cai, ``Gpu-accelerated dislocation
  dynamics using subcycling time-integration,'' {\em Modelling and Simulation
  in Materials Science and Engineering}, vol.~27, no.~7, p.~075014, 2019.

\bibitem{stricker2015dislocation}
M.~Stricker and D.~Weygand, ``Dislocation multiplication mechanisms--glissile
  junctions and their role on the plastic deformation at the microscale,'' {\em
  Acta Materialia}, vol.~99, pp.~130--139, 2015.

\bibitem{sudmanns2019dislocation}
M.~Sudmanns, M.~Stricker, D.~Weygand, T.~Hochrainer, and K.~Schulz,
  ``Dislocation multiplication by cross-slip and glissile reaction in a
  dislocation based continuum formulation of crystal plasticity,'' {\em Journal
  of the Mechanics and Physics of Solids}, vol.~132, p.~103695, 2019.

\bibitem{kocks2003physics}
U.~Kocks and H.~Mecking, ``Physics and phenomenology of strain hardening: the
  fcc case,'' {\em Progress in materials science}, vol.~48, no.~3,
  pp.~171--273, 2003.

\bibitem{devincre2008dislocation}
B.~Devincre, T.~Hoc, and L.~Kubin, ``Dislocation mean free paths and strain
  hardening of crystals,'' {\em Science}, vol.~320, no.~5884, pp.~1745--1748,
  2008.

\bibitem{demir2016physically}
E.~Demir, ``A physically based constitutive model for fcc single crystals with
  a single state variable per slip system,'' {\em Modelling and Simulation in
  Materials Science and Engineering}, vol.~25, no.~1, p.~015009, 2016.

\bibitem{kubin2008modeling}
L.~Kubin, B.~Devincre, and T.~Hoc, ``Modeling dislocation storage rates and
  mean free paths in face-centered cubic crystals,'' {\em Acta materialia},
  vol.~56, no.~20, pp.~6040--6049, 2008.

\bibitem{csikor2005numerical}
F.~Csikor, B.~Kocsis, B.~Bak{\'o}, and I.~Groma, ``Numerical characterisation
  of the relaxation of dislocation systems,'' {\em Materials Science and
  Engineering: A}, vol.~400, pp.~214--217, 2005.

\bibitem{honeycomb1972plastic}
R.~Honeycomb, ``Plastic deformation of metals,'' 1972.

\bibitem{takeuchi1975work}
T.~Takeuchi, ``Work hardening of copper single crystals with multiple glide
  orientations,'' {\em Transactions of the Japan Institute of Metals}, vol.~16,
  no.~10, pp.~629--640, 1975.

\bibitem{taylor1934mechanism}
G.~I. Taylor, ``The mechanism of plastic deformation of crystals. part
  i.—theoretical,'' {\em Proceedings of the Royal Society of London. Series
  A, Containing Papers of a Mathematical and Physical Character}, vol.~145,
  no.~855, pp.~362--387, 1934.

\bibitem{neuhaus1992flow}
R.~Neuhaus and C.~Schwink, ``On the flow stress of [100]-and [111]-oriented
  cu-mn single crystals: A transmission electron microscopy study,'' {\em
  Philosophical Magazine A}, vol.~65, no.~6, pp.~1463--1484, 1992.

\bibitem{madec2002dislocation}
R.~Madec, B.~Devincre, and L.~Kubin, ``From dislocation junctions to forest
  hardening,'' {\em Physical review letters}, vol.~89, no.~25, p.~255508, 2002.

\bibitem{mecking1981kinetics}
H.~Mecking and U.~Kocks, ``Kinetics of flow and strain-hardening,'' {\em Acta
  Metallurgica}, vol.~29, no.~11, pp.~1865--1875, 1981.

\bibitem{edington1969influence}
J.~Edington, ``The influence of strain rate on the mechanical properties and
  dislocation substructure in deformed copper single crystals,'' {\em
  Philosophical Magazine}, vol.~19, no.~162, pp.~1189--1206, 1969.

\bibitem{akhondzadeh2018geometrically}
S.~Akhondzadeh, R.~Sills, S.~Papanikolaou, E.~Van~der Giessen, and W.~Cai,
  ``Geometrically projected discrete dislocation dynamics,'' {\em Modelling and
  Simulation in Materials Science and Engineering}, vol.~26, no.~6, p.~065011,
  2018.

\bibitem{franciosi1982multislip}
P.~Franciosi and A.~Zaoui, ``Multislip tests on copper crystals: a junctions
  hardening effect,'' {\em Acta Metallurgica}, vol.~30, no.~12, pp.~2141--2151,
  1982.

\bibitem{madec2003role}
R.~Madec, B.~Devincre, L.~Kubin, T.~Hoc, and D.~Rodney, ``The role of collinear
  interaction in dislocation-induced hardening,'' {\em Science}, vol.~301,
  no.~5641, pp.~1879--1882, 2003.

\bibitem{devincre2006physical}
B.~Devincre, L.~Kubin, and T.~Hoc, ``Physical analyses of crystal plasticity by
  dd simulations,'' {\em Scripta Materialia}, vol.~54, no.~5, pp.~741--746,
  2006.

\bibitem{messner2017crystal}
M.~C. Messner, M.~Rhee, A.~Arsenlis, and N.~R. Barton, ``A crystal plasticity
  model for slip in hexagonal close packed metals based on discrete dislocation
  simulations,'' {\em Modelling and Simulation in Materials Science and
  Engineering}, vol.~25, no.~4, p.~044001, 2017.

\bibitem{sills2018dislocation}
R.~B. Sills, N.~Bertin, A.~Aghaei, and W.~Cai, ``Dislocation networks and the
  microstructural origin of strain hardening,'' {\em Physical review letters},
  vol.~121, no.~8, p.~085501, 2018.

\bibitem{saada1960hardening}
G.~Saada, ``On hardening due to the recombination of dislocations,'' {\em Acta
  Metall}, vol.~8, pp.~841--847, 1960.

\bibitem{stricker2018dislocation}
M.~Stricker, M.~Sudmanns, K.~Schulz, T.~Hochrainer, and D.~Weygand,
  ``Dislocation multiplication in stage ii deformation of fcc multi-slip single
  crystals,'' {\em Journal of the Mechanics and Physics of Solids}, vol.~119,
  pp.~319--333, 2018.

\bibitem{steinberg1988constitutive}
D.~Steinberg and C.~Lund, ``A constitutive model for strain rates from 10-4 to
  106 s-1,'' {\em Le Journal de Physique Colloques}, vol.~49, no.~C3,
  pp.~C3--433, 1988.

\bibitem{frost1971motion}
H.~Frost and M.~Ashby, ``Motion of a dislocation acted on by a viscous drag
  through an array of discrete obstacles,'' {\em Journal of Applied Physics},
  vol.~42, no.~13, pp.~5273--5279, 1971.

\bibitem{kocks1975thermodynamics}
U.~Kocks, A.~Argon, and M.~Ashby, ``Thermodynamics and kinetics of slip,'' {\em
  Progress in Material Science}, vol.~19, pp.~1--291, 1975.

\bibitem{busso1996dislocation}
E.~P. Busso and F.~A. McClintock, ``A dislocation mechanics-based
  crystallographic model of a b2-type intermetallic alloy,'' {\em International
  Journal of Plasticity}, vol.~12, no.~1, pp.~1--28, 1996.

\bibitem{mecking1970new}
H.~Mecking and K.~L{\"u}cke, ``A new aspect of the theory of flow stress of
  metals,'' {\em Scripta Metallurgica}, vol.~4, no.~6, pp.~427--432, 1970.

\bibitem{essmann1979annihilation}
U.~Essmann and H.~Mughrabi, ``Annihilation of dislocations during tensile and
  cyclic deformation and limits of dislocation densities,'' {\em Philosophical
  Magazine A}, vol.~40, no.~6, pp.~731--756, 1979.

\bibitem{casals2007crystal}
O.~Casals, J.~O{\v{c}}en{\'a}{\v{s}}ek, and J.~Alcala, ``Crystal plasticity
  finite element simulations of pyramidal indentation in copper single
  crystals,'' {\em Acta materialia}, vol.~55, no.~1, pp.~55--68, 2007.

\bibitem{weygand2014mechanics}
D.~Weygand, ``Mechanics and dislocation structures at the micro-scale: Insights
  on dislocation multiplication mechanisms from discrete dislocation dynamics
  simulations,'' {\em MRS Online Proceedings Library Archive}, vol.~1651, 2014.

\bibitem{zepedaruiz2019metal}
L.~A. Zepeda-Ruiz, A.~Stukowski, T.~Oppelstrup, N.~Bertin, N.~R. Barton,
  R.~Freitas, and V.~V. Bulatov, ``Metal hardening in atomistic detail,'' 2019.

\bibitem{ScikitGP}
``Scikit-learn.''
\newblock https://scikit-learn.org/stable/modules/gaussian\_process.html.

\bibitem{rasmussen2003gaussian}
C.~E. Rasmussen, ``Gaussian processes in machine learning,'' in {\em Summer
  School on Machine Learning}, pp.~63--71, Springer, 2003.

\bibitem{molinari1987self}
A.~Molinari, G.~Canova, and S.~Ahzi, ``A self consistent approach of the large
  deformation polycrystal viscoplasticity,'' {\em Acta Metallurgica}, vol.~35,
  no.~12, pp.~2983--2994, 1987.

\bibitem{lebensohn1993self}
R.~A. Lebensohn and C.~Tom{\'e}, ``A self-consistent anisotropic approach for
  the simulation of plastic deformation and texture development of
  polycrystals: application to zirconium alloys,'' {\em Acta metallurgica et
  materialia}, vol.~41, no.~9, pp.~2611--2624, 1993.

\bibitem{higashida1986formation}
K.~Higashida, J.-i. Takamura, and N.~Narita, ``The formation of deformation
  bands in fcc crystals,'' {\em Materials science and engineering}, vol.~81,
  pp.~239--258, 1986.

\bibitem{kang2014stress}
K.~Kang, J.~Yin, and W.~Cai, ``Stress dependence of cross slip energy barrier
  for face-centered cubic nickel,'' {\em Journal of the Mechanics and Physics
  of Solids}, vol.~62, pp.~181--193, 2014.

\bibitem{hussein2015microstructurally}
A.~M. Hussein, S.~I. Rao, M.~D. Uchic, D.~M. Dimiduk, and J.~A. El-Awady,
  ``Microstructurally based cross-slip mechanisms and their effects on
  dislocation microstructure evolution in fcc crystals,'' {\em Acta
  Materialia}, vol.~85, pp.~180--190, 2015.

\bibitem{shanthraj2011dislocation}
P.~Shanthraj and M.~Zikry, ``Dislocation density evolution and interactions in
  crystalline materials,'' {\em Acta materialia}, vol.~59, no.~20,
  pp.~7695--7702, 2011.

\bibitem{fivel1998identification}
M.~Fivel, L.~Tabourot, E.~Rauch, and G.~Canova, ``Identification through
  mesoscopic simulations of macroscopic parameters of physically based
  constitutive equations for the plastic behaviour of fcc single crystats,''
  {\em Le Journal de Physique IV}, vol.~8, no.~PR8, pp.~Pr8--151, 1998.

\bibitem{devincre2015physically}
B.~Devincre and R.~Gatti, ``Physically justified models for crystal plasticity
  developed with dislocation dynamics simulations,'' 2015.

\end{thebibliography}


\end{document}